\documentclass[12pt]{article}
\usepackage{amsfonts}
\usepackage{amssymb}
\usepackage{amsbsy}
\usepackage{graphics}
\usepackage{epsfig}
\usepackage{color}

\input epsf.sty

\newlength{\TZ}
\newcommand{\BEQ}{\begin{equation}}     
\newcommand{\BEA}{\begin{eqnarray}}
\newcommand{\BD}{\begin{displaymath}}
\newcommand{\EEQ}{\end{equation}}       
\newcommand{\EEA}{\end{eqnarray}}
\newcommand{\ED}{\end{displaymath}}
\newcommand{\bb}{\begin{eqnarray}}
\newcommand{\ee}{\end{eqnarray}}

\newcommand{\e}{{\rm e}}

\newcommand{\D}{{\rm d}}                
\newcommand{\II}{{\rm i}}               
 
\newcommand{\demi}{\frac{1}{2}}         
\newcommand{\tr}{{\rm tr\,}}            
\newcommand{\wit}[1]{\widetilde{#1}}    
\newcommand{\wht}[1]{\widehat{#1}}      

\renewcommand{\vec}[1]{\boldsymbol{#1}} 

\newcommand{\appsection}[2]{\setcounter{equation}{0}\setcounter{subsection}{0}
\section*{Appendix #1. #2}
\renewcommand{\theequation}{#1.\arabic{equation}}
              \renewcommand{\thesection}{#1} }

\catcode`\@=11
\def\numberbysection{\@addtoreset{equation}{section}
        \def\theequation{\thesection.\arabic{equation}}}
\numberbysection

\begin{document}

\begin{titlepage}

\vskip 1.5 cm
\begin{center}
{\Large \bf Quantum phase transition in the spin-anisotropic quantum spherical model}
\end{center}

\vskip 2.0 cm
\centerline{{\bf Sascha Wald} and {\bf Malte Henkel}}
\vskip 0.5 cm
\begin{center}
Groupe de Physique Statistique, \\
D\'epartement de Physique de la Mati\`ere et des Mat\'eriaux, \\
Institut Jean Lamour (CNRS UMR 7198),  Universit\'e de Lorraine Nancy, \\ 
B.P. 70239, F -- 54506 Vand{\oe}uvre l\`es Nancy Cedex, France
\end{center}

\begin{abstract}
Motivated by an analogy with the spin anisotropies in the quantum XY chain and its reformulation in terms of
spin-less Majorana fermions, its bosonic analogue, the spin-anisotropic quantum spherical model, is introduced. 
The exact solution of the model permits to analyse the influence of the spin-anisotropy on the phase diagram 
and the universality of the critical behaviour in a new way, since the interactions of the quantum spins and their conjugate
momenta create new effects. At zero temperature, a quantum critical line is found, which is in the same universality
class as the thermal phase transition in the classical spherical model in $d+1$ dimensions. The location of this
quantum critical line shows a re-entrant quantum phase transition for dimensions $1<d\lesssim 2.065$. 
\end{abstract}

\vfill
~\\

\end{titlepage}

\setcounter{footnote}{0} 

\section{Introduction}

The study of equilibrium phase transitions has taken enormous benefits from the analysis of
exactly solvable models \cite{Joyc72,Baxt82,Henk99,Sach99,Bran00,Camp14b}. 
The classical spherical model, invented in the seminal work by Berlin and Kac \cite{Berl52} 
and with its subsequent simplification by Lewis and Wannier \cite{Lewi52}, 
has been a valuable test system  for the explicit analytical verification of more general 
scaling descriptions, in a specific setting (examples are critical behaviour of observables or finite-size scaling). 
It is related with more realistic spin systems as the $n\to\infty$ 
limit of the ${\rm O}(n)$-symmetric Heisenberg model \cite{Stan68}. 
It is well-known, as already observed by Berlin and Kac \cite{Berl52}, 
that in the original formulation in terms of classical spin variables
$S_i\in\mathbb{R}$, the specific heat does not vanish in the 
zero-temperature limit, hence the Nernst theorem is not obeyed in this model. 
This was one motivation to take the quantum nature of the spin variables 
into account, by a canonical quantisation scheme, and has lead to Obermair's formulation of the
{\em quantum spherical model} \cite{Ober72}. This takes the form of a quantum rotor model, 
where the kinetic energy term in the hamiltonian does not 
commute with the spin-exchange interactions. The properties of
this exactly solvable model have been analysed in great detail, 
see e.g. \cite{Ober72,Shuk81,Niew95,Vojt96,Cham98,Sach99,Bran00,Grac04,Oliv06,Bien12,Bien13}. 
Independently, the quantum spherical model was also 
obtained via the so-called `hamiltonian limit' \cite{Kogu79} as the logarithm of the
transfer matrix in an extremely anisotropic limit \cite{Sred79,Henk84a,Henk92}. In its most conventional formulation as a
quantum rotor model \cite{Ober72,Henk84a,Vojt96}, the quantum spherical model may be obtained as the limit $n\to\infty$ 
of the quantum non-linear O($n$) sigma-model \cite{Vojt96}. For different choices of the kinetic energy term, there are further 
quantum spherical models, which become the $n\to\infty$ limit of an
SU($n$) Heisenberg ferromagnet or anti-ferromagnet \cite{Niew95,Grac04}. 

Quantum spherical models have been discussed in the context of specific applications, for example for the description of networks
of Josephson-junction arrays \cite{Fazi01,Cha03}. Certain modern theories of cuprate supraconductivity are based on 
SO(5)-symmetric quantum non-linear sigma models, and it is thought that this kind of models might be an effective
description of the large-distance, low-energy properties of more realistic models, see e.g. \cite{Deml04} for a detailed review.

Habitually, (mean) spherical models are defined in terms of a classical hamiltonian
\BEQ
{\cal H}_{\rm cl} = \sum_{\vec{n}} \left[ -J \sum_{j=1}^d {S}_{\vec{n}}{S}_{\vec{n}+\vec{e}_j} - B S_{\vec{n}} +\frac{\mu}{2} {S}_{\vec{n}}^2 \right]
\EEQ
with the spherical spins ${S}_{\vec{n}}\in\mathbb{R}$.  
Herein, $\vec{n}$ runs over the sites of a $d$-dimensional hyper-cubic lattice with ${\cal N}=N^d$ sites, 
the vectors $\vec{e}_j$, $1\leq j \leq d $, are the unit vectors in the $j^{\rm th}$ direction, 
$B$ is an external magnetic field  and $J$ is the exchange integral. 
Finally, the spherical spins obey the mean `spherical constraint' 
$\sum_{\vec{n}} \left\langle S_{\vec{n}}^2\right\rangle \stackrel{!}{=}{\cal N}$ \cite{Berl52,Lewi52}, from 
which $\mu$ is found.

The generalisation towards a {\em quantum spherical model} is formulated by considering now the spins 
$S_{\vec{n}}\mapsto \wht{S}_{\vec{n}}$ as operators, 
and introducing canonically conjugate momenta $\wht{P}_{\vec{n}}$, which obey the canonical commutation relations
\BEQ 
\left[ \wht{S}_{\vec{n}}, \wht{P}_{\vec{m}} \right] = \II \hbar\, \delta_{\vec{n},\vec{m}} \;\; , \;\;
\left[ \wht{S}_{\vec{n}}, \wht{S}_{\vec{m}} \right] = \left[ \wht{P}_{\vec{n}}, \wht{P}_{\vec{m}} \right] = 0
\EEQ
The most common ansatz for the quantum hamiltonian is to make 
${\cal H}_{\rm cl}\mapsto \wht{H}_{\rm cl}$ an operator and to add a kinetic energy
term of non-interacting momenta,\footnote{Even in the case of competing interactions, where
new multicritical points, called {\em Lifshitz points} \cite{Horn75}, can be found in the classical spherical model and which may present strongly
anisotropic scaling behaviour, see \cite{Folk93,Frac93,Dieh03,Hase06,Henk10,Shpo12} and refs. therein, 
existing studies on the quantum version do not consider any interactions between the momenta \cite{Gome13}.} viz. 
\BEQ \label{gl:Hini}
\wht{H} = \wht{H}_{\rm cl} + \frac{g}{2} \sum_{\vec{n}}  \wht{P}_{\vec{n}}^2 
= \sum_{\vec{n}} \left[  -J \sum_{j=1}^d \wht{S}_{\vec{n}}\wht{S}_{\vec{n}+\vec{e}_j} 
- B \wht{S}_{\vec{n}}+\frac{\mu}{2} \wht{S}_{\vec{n}}^2 + \frac{g}{2} \wht{P}_{\vec{n}}^2 \right]
\EEQ
with a new coupling $g$, which controls the strength of the quantum fluctuations.  
In equilibrium, one can express the spherical constraint\footnote{Sometimes the constraint is given in the form 
$\sum_{\vec{n}} \left\langle \wht{S}_{\vec{n}}^2\right\rangle = {\cal N}/4$, see e.g. \cite{Vojt96,Oliv06},
which in the zero temperature limit amounts essentially to a re-scaling of the spherical parameter. 
Throughout, units are such that the Boltzmann constant $k_B=1$.}  
as a thermodynamic derivative
\BEQ \label{gl:cs}
\sum_{\vec{n}} \left\langle \wht{S}_{\vec{n}}^{\,2} \right\rangle = - \frac{2}{T} \frac{\partial \ln \cal Z}{\partial \mu} \stackrel{!}{=} {\cal N}
\EEQ
where ${\cal Z}=\tr \exp(-\wht{H}/T)$ is the partition function and $T$ is the temperature. 
This quantum hamiltonian can also be obtained as the logarithm of the transfer matrix of the 
classical spherical model in $d+1$ dimensions, in a certain strongly anisotropic limit
\cite{Suzu71,Kogu79,Henk84a,Henk99,Sach99}. This mapping in particular shows 
that the zero-temperature quantum critical behaviour of the quantum phase transition of the ground-state of 
the quantum spherical model (\ref{gl:Hini}) in $d$ dimensions  \cite{Henk84a} is in the 
same universality class as the finite-temperature transition of the
classical spherical model in $d+1$ dimensions \cite{Niew95,Vojt96,Sach99,Bran00,Grac04,Oliv06}. 

It is straightforward to recast the hamiltonian (\ref{gl:Hini})
in terms of bosonic ladder operators $\wht{a}_{\vec{n}}$ and $\wht{a}_{\vec{n}}^{\dag}$, defined as follows \cite{Ober72}
\BEQ \label{1.5}
\wht{S}_{\vec{n}} = \sqrt{\frac{\hbar}{2}\,}\,\left(\frac{g}{\mu}\right)^{1/4} 
\left( \wht{a}_{\vec{n}} + \wht{a}_{\vec{n}}^{\dag} \right) 
\;\; , \;\;
\wht{P}_{\vec{n}} = \frac{1}{\II}\sqrt{\frac{\hbar}{2}\,}\,\left(\frac{\mu}{g}\right)^{1/4} 
\left( \wht{a}_{\vec{n}} - \wht{a}_{\vec{n}}^{\dag} \right) 
\EEQ
which obey the canonical commutator relations
\BEQ
\left[ \wht{a}_{\vec{n}}, \wht{a}_{\vec{m}}^{\dag} \right] = \delta_{\vec{n},\vec{m}} \;\; , \;\;
\left[ \wht{a}_{\vec{n}}, \wht{a}_{\vec{m}} \right] =\left[ \wht{a}_{\vec{n}}^{\dag}, \wht{a}_{\vec{m}}^{\dag} \right] = 0
\EEQ
and render the hamiltonian (\ref{gl:Hini}) as follows
\BEA
H &=& \sum_{\vec{n}} \left[  \hbar\sqrt{g\mu\,}\,\left( \wht{a}_{\vec{n}}^{\dag}\wht{a}_{\vec{n}} +\demi\right) 
- B \sqrt{\frac{\hbar}{2}\,}\,\left(\frac{g}{\mu}\right)^{1/4}\left( \wht{a}_{\vec{n}} + \wht{a}_{\vec{n}}^{\dag} \right) \right.
\nonumber \\
& & \left. - \frac{J\hbar}{2}\sqrt{\frac{g}{\mu}\,}\, \sum_{j=1}^d 
\left(  
\wht{a}_{\vec{n}}^{\dag}\wht{a}_{\vec{n}+\vec{e}_j} +\wht{a}_{\vec{n}}\wht{a}_{\vec{n}+\vec{e}_j}^{\dag}  
+ \wht{a}_{\vec{n}}^{\dag}\wht{a}_{\vec{n}+\vec{e}_j}^{\dag} + \wht{a}_{\vec{n}}\wht{a}_{\vec{n}+\vec{e}_j}  
\right) \right]
\label{gl:Hsm1}
\EEA
The computation of the eigenvalues of such hamiltonians is a matter of finding 
the appropriate canonical transformation and is treated in appendix~A. 
Here, we wish to point out an analogy with quantum Ising/XY chains
(also called Ising/XY chains in a transverse field), with an anisotropy in spin space, and given by the hamiltonian \cite{Kats62,Bar71} 
\BEA 
H_{\rm XY} &=& -\frac{1}{2} \sum_n \left[ g \sigma_n^z + \frac{1+\lambda}{2} \sigma_n^x \sigma_{n+1}^x 
+ \frac{1-\lambda}{2} \sigma_n^y \sigma_{n+1}^y \right] 
\label{gl:IXY} \\
&=&  \sum_n \left[ g \left( \wht{c}_n^{\dag} \wht{c}_n - \demi\right) 
-\demi \left( \wht{c}_n^{\dag} \wht{c}_{n+1} - \wht{c}_n \wht{c}_{n+1}^{\dag} 
+\lambda\left( \wht{c}_n^{\dag} \wht{c}_{n+1}^{\dag} - \wht{c}_n \wht{c}_{n+1} \right)\right) \right]
\label{gl:IXY_ferm}
\EEA
where the $\sigma_n^{x,y,z}$ denote the Pauli matrices 
attached to the $n^{\rm th}$ site of a periodic chain of $N$ sites. 
The transverse field $g$ measures the quantum fluctuations and $\lambda$ is a spin-anisotropy coupling. 
After a Jordan-Wigner transformation, the hamiltonian (\ref{gl:IXY}) 
can be brought to a quadratic form (\ref{gl:IXY_ferm}) in the fermionic ladder operators $\wht{c}_n$ and $\wht{c}_n^\dagger$
(we did not carefully specify the non-local boundary conditions 
in the fermionic variables since we shall not require their form) with the anticommutator relations
\BEQ
\left\{ \wht{c}_n, \wht{c}_m^{\dag} \right\} = \delta_{n,m} \;\; , \;\; 
\left\{\wht{c}_n, \wht{c}_m \right\} =  \left\{ \wht{c}_n^{\dag}, \wht{c}_m^{\dag} \right\} = 0
\EEQ
The ground-state of quantum Ising/XY chain (\ref{gl:IXY}) has a rich phase 
diagram with a disordered phase for $g>1$, a line of second-order
transitions at $g=1$ which is in the universality class of the $2D$ Ising model for $\lambda\ne0$, 
an ordered ferromagnetic phase for $\sqrt{1-\lambda^2}<g<1$ and an ordered oscillating phase for 
$g<\sqrt{1-\lambda^2}$ \cite{Bar71,Henk87,Chak96,Henk99,Kare00,Dutt10}. The universality of the quantum critical behaviour at $T=0$, including
the universal amplitude combinations \cite{Priv84,Priv91,Henk01,Camp14}, with respect to $0<\lambda\leq 1$ along
the Ising critical line has been explicitly confirmed: for the chain for both the spin-$\demi$ as well as the the spin-$1$ representations of
the Lie algebra of the rotation group \cite{Henk87}, as well as in $2D$ for the  spin-$\demi$ representation \cite{Henk8487}. 

Comparing the fermionic hamiltonian (\ref{gl:IXY_ferm}) with the bosonic one (\ref{gl:Hsm1}),\footnote{Alternatively, one can
consider the fermionic degrees of freedom in (\ref{gl:IXY_ferm}) as {\em hard-core bosons}. Relaxing the `hard-core/fermionic'
constraint on the single-site occupation numbers $\langle \wht{n}_i\rangle = \langle\wht{c}_i^{\dag} \wht{c}_i\rangle \stackrel{!}{=}0,1$, 
towards $\sum_i \langle\wht{c}_i^{\dag} \wht{c}_i\rangle \stackrel{!}{=} \bar{\nu}{\cal N}$, 
where $\bar{\nu}=\demi$ is a filling factor, one has a third way to replace (\ref{gl:IXY_ferm}) by a quantum spherical model \cite{Ma97}.} 
one observes that in the former the two-particle
annihilation/creation processes are controlled by the parameter $\lambda$, 
whereas that parameter happens to be fixed to unity in the latter. 
Here, we shall inquire into what happens if an analogous rate is introduced 
into the hamiltonian (\ref{gl:Hsm1}), and write 
\BEA
H &=& 
\sqrt{g\mu\hbar^2\,}\: \sum_{\vec{n}} \left[  \left( \wht{a}_{\vec{n}}^{\dag}\wht{a}_{\vec{n}} +\demi\right) 
- B \left(\frac{1}{4\hbar^2 g\mu^3}\right)^{1/4}\left( \wht{a}_{\vec{n}} + \wht{a}_{\vec{n}}^{\dag} \right) \right.
\nonumber \\
& & \left. - \frac{J}{\mu} \sum_{j=1}^d 
\left(  
\wht{a}_{\vec{n}}^{\dag}\wht{a}_{\vec{n}+\vec{e}_j} +\wht{a}_{\vec{n}}\wht{a}_{\vec{n}+\vec{e}_j}^{\dag}  
+ \lambda \left( \wht{a}_{\vec{n}}^{\dag}\wht{a}_{\vec{n}+\vec{e}_j}^{\dag} + \wht{a}_{\vec{n}}\wht{a}_{\vec{n}+\vec{e}_j} \right) 
\right) \right]
\label{gl:Hsm2}
\\
&=& \sum_{\vec{n}} \left[ \frac{g}{2} \wht{P}_{\vec{n}}^2 + \frac{\mu}{2} \wht{S}_{\vec{n}}^2 - B \wht{S}_{\vec{n}} 
-\frac{1}{4s}\sum_{j=1}^d \left( (1+\lambda) \mu\, \wht{S}_{\vec{n}}\wht{S}_{\vec{n}+\vec{e}_j} 
+ (1-\lambda) g\, \wht{P}_{\vec{n}}\wht{P}_{\vec{n}+\vec{e}_j} \right)\right]
\nonumber
\EEA
The re-formulation in terms of the original spins and momenta 
shows that the hamiltonian (\ref{gl:Hsm2}) introduces an interaction between the momenta, 
quite analogous to the spin anisotropies in the quantum XY chain (\ref{gl:IXY}). 
In the special case $\lambda=1$, this new interaction disappears and one is back to the 
quantum rotor spherical model as studied in the literature so far. We call the model defined by (\ref{gl:Hsm2}) the
{\em spin-anisotropic quantum spherical model} ({\sc saqsm}), 
because of the analogy of the parameter $\lambda$ with the spin anisotropy in the
fermionic hamiltonian (\ref{gl:IXY},\ref{gl:IXY_ferm}). 

It will be
convenient to work with the spherical parameter (already used in (\ref{gl:Hsm2})) 
\BEQ
s := \frac{\mu}{2 J}
\EEQ
For $B=0$, there is a duality transformation 
$\wht{S}_{\vec{n}} \leftrightarrow \wht{P}_{\vec{n}}$, 
$\mu \leftrightarrow g$, $\lambda \leftrightarrow -\lambda$. 
It is therefore sufficient to restrict attention to the case $\lambda\geq 0$, as we shall do from now on.  
In the special case $\lambda=0$, pairs of particles can neither be created, 
nor destroyed, which formally is expressed through the 
conservation, expressed by $[\wht{N},H]=0$, of  the total number of particles
$\wht{N} := \sum_{\vec{n}} \wht{a}_{\vec{n}}^{\dag} \wht{a}_{\vec{n}}$. 
This case has properties different from the situation where $\lambda\ne 0$.\footnote{The conservation of $\wht{N}$ is reminiscent of
the spherical constraints used in \cite{Niew95,Grac04}, although the quantum critical behaviour of the $\lambda=0$ model (\ref{gl:Hsm2})
will turn out to be different.}

This work is organised as follows. Section~2 presents the general formalism for the solution of the model and the new techniques required
for its analysis when $\lambda\ne 0,1$. We shall focus on the quantum phase transition at zero temperature. A detailed analysis of the spherical constraint
surprisingly shows that for dimensions $1<d\lesssim 2.065$, there is a re-entrant quantum phase transitions when $\lambda$ is small enough. 
There is no known classical analogue of this effect. The critical behaviour and its universality along the $\lambda$-dependent critical
lines will be analysed and we shall discuss the relationship with the thermal phase transition of the classical spherical model. 
As one should have expected, we find a critical line\footnote{Our methods of analysis are restricted to $|\lambda|\leq 1$, see appendices~B and~C.} 
for $0<\lambda\leq 1$, where the quantum critical behaviour of the {\sc saqsm} is in the
same universality class as in the classical spherical model in $d+1$ dimensions. 
Section~3 gives our conclusions. Technical details are treated in several appendices. Appendix~A recalls the 
exact diagonalisation techniques, in appendices~B and~C the spherical constraint and the consequences for the quantum critical point are studied,
in appendix~D the spin-spin correlator is derived and appendix~E looks in more detail into the existence of the re-entrant quantum phase transition. 

\section{Solution and quantum phase transition}

\subsection{General formalism} 

In order to analyse the thermodynamic behaviour of the quantum spherical model (\ref{gl:Hsm2}) with $\lambda$ arbitrary, 
the first task is to bring $H$ into a diagonal form. This calculation is carried out in appendix~A, and leads to
\BEQ \label{eq:H}
H =\sqrt{2\hbar^2gJ/s} \sum_{\vec{k} \in {\cal{K}}} \Lambda_{\vec{k}} 
\left( \wht{b}_{\vec{k}}^{\dag} \wht{b}_{\vec{k}} + \demi \right) + H_0
\EEQ
where the eigenvalues are given in eq.~(\ref{A9}) 
\BEQ 
\Lambda_{\vec{k}} := s\bar{\Lambda}_{\vec{k}}= \sqrt{ s - \frac{1+\lambda}{2} \sum_{j=1}^d \cos k_j} \: 
\sqrt{ s - \frac{1-\lambda}{2} \sum_{j=1}^d \cos k_j}
\EEQ
and the quasi-momenta ${\cal{K}} \ni k_j= \frac{2\pi}{N} n_j$, with $n_j=0,1,\ldots N-1$ 
and $j=1,\ldots,d$, with the reciprocal lattice ${\cal{K}}$. Finally, from (\ref{A16}) we have
\BEQ
H_0 = \frac{B^2}{4J} \frac{{\cal N}}{s-(1+\lambda)d/(2)} 
\EEQ
Since the quasi-particles are independent, non-interacting particles, 
the calculation of the partition function reduces to a computation
of products of geometric series, such that the 
free energy $F=-T\ln {\cal Z}$ reads explicitly
\BEA
\lefteqn{F = T {\cal N} \ln 2 - \frac{B^2}{4J} \frac{{\cal N}}{s - (1+\lambda)d/2}} 
\nonumber \\ \label{gibbs}
&+& T \sum_{\vec{k}} \ln \sinh \left[ \frac{\hbar}{T} \sqrt{\frac{g J}{2s}\,}\, 
\sqrt{ s - \frac{1+\lambda}{2} \sum_{j=1}^d \cos k_j\,}\:
\sqrt{ s - \frac{1-\lambda}{2} \sum_{j=1}^d \cos k_j\,}\;\right]
\EEA
At this point, one can go to the infinite-size limit ${\cal N}=N^d\to\infty$. 
In particular, the spherical constraint (\ref{gl:cs}) then takes the form
\BEQ \label{gl:2.5}
\sqrt{\frac{g \hbar^2}{2J s}} \int_{{\cal B}} \frac{\D\vec{k}}{(2\pi)^d} 
\coth\left(\sqrt{\frac{g J\hbar^2}{2T^2 s}} \Lambda_{\vec{k}} \right)
\frac{s - \frac{1-\lambda^2}{4s} \left[\sum_{j=1}^d \cos k_j\right]^2}{2\Lambda_{\vec{k}}} 
+\left( \frac{B}{2J}\right)^2\left(s-\frac{1+\lambda}{2}d\right)^{-2} =1
\EEQ
where ${\cal B}=[-\pi,\pi]^d$ is the Brillouin zone. 
For the special case $\lambda=1$, we recover the form of the spherical constraint 
known from the literature, see \cite{Henk84a,Vojt96,Sach99,Bran00,Oliv06}. 

Besides thermodynamic observables, we shall also study the spin-spin correlator. In appendix~D, it is shown that
\BEQ \label{2.6}
\left< S_{\vec{n}} S_{\vec{n}+\vec{r}}\right>=\sqrt{\frac{\hbar^2g}{8Js}} 
\int_{\cal{B}}\frac{\D \vec{k}}{(2\pi)^d}\sqrt{\frac{2s-(1-\lambda)\sum_{j=1}^d \cos k_j}{2s-(1+\lambda)\sum_{j=1}^d \cos k_j}}
\coth \left[\sqrt{2\hbar^2gJ/s}\Lambda_{\vec{k}}/(2T) \right] \prod_{j=1}^d \cos \left(r_j k_j\right)
\EEQ

\subsection{Quantum phase transition}

In $d>2$ dimensions, the spherical model undergoes a phase transition 
at some critical temperature $T_c>0$ \cite{Vojt96,Ma97,Sach99,Bran00,Grac04,Oliv06}. 
In general, one expects that this finite-temperature transition of the 
$d$-dimensional model should be in the same universality class 
as the one of the classical model (without quantum terms) \cite{Kogu79,Sach99,Bran00}. 
Here, we rather concentrate on the quantum phase transition which occurs in the
ground-state, that is, at temperature $T=0$. 

Generically, quantum phase transitions arise mathematically from a degeneracy in the ground-state of the hamiltonian. 
In order to localise the quantum critical point in terms of the model's parameters, 
consider the smallest energy gap $\Delta E$
\BEQ \label{gap}
\Delta E := \lim_{\vec{k}\to\vec{0}} \Lambda_{\vec{k}} 
= \sqrt{ s - \frac{1+\lambda}{2} d\:}\:\sqrt{ s - \frac{1-\lambda}{2} d\:} 
\EEQ
This energy gap closes for 
\BEQ \label{gl:scrit}
s_c := \frac{1+|\lambda|}{2} d
\EEQ
such that the spherical parameter must satisfy $s\geq (1+|\lambda|)d/2$. 
The ground-state thermodynamics now follows from 
an analysis of the spherical constraint (\ref{gl:2.5}), which in the limit $T\to 0$ takes the form
\BEQ \label{2.8}
\sqrt{\frac{g \hbar^2}{8J s^3}} \int_{\cal B}\frac{\D \vec{k}}{(2\pi)^d} 
\frac{s^2 -\frac{1-\lambda^2}{4}\left[ \sum_{j=1}^d \cos k_j\right]^2}
{[s-\frac{1+\lambda}{2}\sum_{j=1}^d\cos k_j]^{1/2}\,[s-\frac{1-\lambda}{2}\sum_{j=1}^d\cos k_j]^{1/2}}
+\left(\frac{B}{2J}\frac{1}{s-\frac{1+\lambda}{2}d}\right)^2 =1
\EEQ
This defines the function $s=s(g,\lambda,d,B)$, or alternatively its inverse $g=g(s,\lambda,d,B)$. 
For a vanishing external field $B=0$, this equation is symmetric under $\lambda\mapsto -\lambda$,
hence it is then sufficient to consider the case $\lambda\geq 0$ only. 
We shall almost always restrict to this special case, and then write
$g=g(s,\lambda,d):=g(s,\lambda,d,0)$. 

{\bf 1.} For $\lambda=0$, the constraint simplifies considerably and can be worked out explicitly
\BEQ \label{2.10}
1 = \left( \frac{B}{2J} \frac{1}{s-d/2} \right)^2 + 
\sqrt{\frac{g \hbar^2}{8J s^3}\,} \int_{\cal B} \frac{\D \vec{k}}{(2\pi)^d} \left[ s + \demi \sum_{j=1}^d \cos k_j \right]
=  \left( \frac{B}{2J} \frac{1}{s-d/2} \right)^2 + \sqrt{\frac{g \hbar^2}{8J s}\,} 
\EEQ
Eq.~(\ref{2.10}) gives directly the inverse function $g=g(s,0,d)$, where $d$ appears as a real parameter. 

{\bf 2.} For $\lambda=1$, this has been analysed many times and it is well-known \cite{Henk84a,Vojt96,Cham98,Bran00,Oliv06} 
that (\ref{2.8}) can be re-written as (set $B=0$)
\BEQ \label{2.9}
\sqrt{\frac{2\pi J}{\hbar^2\: g}} = \int_0^{\infty} \!\D u\: e^{-s u^2} I_0(u^2)^d
\EEQ
where $I_0$ is a modified Bessel function \cite{Abra65}. 
Again, this formulation has the appealing feature that by now $d$ can be considered as a continuous parameter in
an analytic continuation $g=g(s,1,d)$. 

{\bf 3.} Finally, for generic $\lambda$, the constraint (\ref{2.8}) can be written in the form (set $B=0$)
\BEA\nonumber
\sqrt{\frac{8\pi^2J}{\hbar^2\: g}} &=& s^{-\frac{3}{2}} \int_0^{\infty} \!\!\D u \int_0^1 \!\D x\: \frac{\exp (-us)}{\sqrt{x(1-x)}\,}\,  I_0(\varrho)^d 
\,\\ 
&\times& 	\left[ s^2 -d(d-1)\frac{1-\lambda^2}{4} \frac{I_1(\varrho)^2}{I_0(\varrho)^2}
- \frac{d}{2}\frac{1-\lambda^2}{4} \left( 1 + \frac{I_2(\varrho)}{I_0(\varrho)} \right) \right]
\label{2.11}
\EEA
where the $I_n$ are modified Bessel functions \cite{Abra65} and we defined the function
\BEQ \label{2.12}
\varrho = \varrho(u,x,\lambda) := u \left( x \frac{1+\lambda}{2} + (1-x) \frac{1-\lambda}{2}\right)
\EEQ
Eqs.~(\ref{2.11},\ref{2.12}) contain $d$ as a real parameter and give directly $g=g(s,\lambda,d)$. 
This form of the constraint is derived in appendix~B.

\subsection{Critical behaviour} 

Now, the constraints (\ref{2.10}), (\ref{2.9}) and (\ref{2.11}) can be used to extract the quantum critical coupling
and the relation between $g$ and the spherical parameter $s$ for the different values of $\lambda$.

{\bf 1.} First, we consider the case $\lambda=0$. From (\ref{2.10}), we have, even for $B\ne 0$
\BEQ
\frac{g}{s} = \frac{8J}{\hbar^2} \left( 1 - \left( \frac{B}{2J}\frac{1}{s-d/2}\right)^2 \right)^2
\EEQ
With the critical value $s_c=d/2$, see eq.~(\ref{gl:scrit}), we have the critical coupling, for $B=0$
\BEQ \label{2.14} 
g_c = g_c(0,d) := g(s_c,0,d) = 4d\, \frac{J}{\hbar^2}
\EEQ
which is non-vanishing for any dimension $d>0$. For the later extraction of the critical exponents, we also note
$(g-g_c)/g_c = (2/d)(s-s_c)$. This linear behaviour is independent of $d$, hence there is no upper critical dimension. 

{\bf 2.} Next, we briefly recall the known result for $\lambda=1$. 
We are interested in finding the critical value $g_c=g_c(1,d):=g(s_c,1,d)$, if it exists and
to obtain the variation of $g$ close to $g_c$, which we can describe in terms of
\BEQ \label{2.15}
t_g := \sqrt{\frac{8 J}{\hbar^2}} \left( \frac{1}{\sqrt{g}} - \frac{1}{\sqrt{g_c}}\right) 
\simeq \sqrt{\frac{8 J}{\hbar^2\,}}\: \frac{g_c-g}{g_c^{3/2}}
\EEQ
Consider $\sigma := s-s_c = s-d$. In order to extract from (\ref{2.9}) 
any non-analytic terms in $\sigma$, one may formally
split \cite{Henk84a} the domain of integration $\int_0^{\infty} = \int_0^{\eta} + \int_{\eta}^{\infty}$. 
The first term, if it exists, will give a analytic
contribution to $g(s)$ near $s\approx s_c$, in particular $g_c=g(s_c)$ in the limit 
$\eta\to\infty$; the second term will give any non-analytic
contributions which may arise. In order to find those, recall the asymptotic form 
$I_0(\rho)\simeq e^{\rho} (2\pi\rho)^{-1/2}$ as $\rho\to\infty$ \cite{Abra65}. 
Then, for $d<3$
\BEQ
t_g \simeq \frac{(2\sqrt{\pi})^{-1}}{(2\pi)^{d/2}} \int_{\eta}^{\infty} \frac{\D u}{u^{d}}\, e^{-\sigma u^2} 
= \frac{\sigma^{(d-1)/2}}{(2\pi)^{(d+1)/2}\sqrt{2}}\int_{\sigma\eta}^{\infty} \frac{\D v\: e^{-v}}{v^{(d-1)/2 +1}}  
\stackrel{\sigma\to 0}{=} \frac{\sigma^{(d-1)/2}}{(2\pi)^{(d+1)/2}}\, \Gamma\left(\frac{1-d}{2}\right)\frac{1}{\sqrt{2}}
\EEQ
(the Gamma function $\Gamma(x)$ \cite{Abra65} is defined via analytic continuation, if needed) 
and only now one also lets $\eta\to\infty$. 
For $d>3$, $t_g\sim \sigma + {\rm O}(\sigma^{(d-1)/2})$ is dominated by the analytic term. Finally, for $d=3$, 
the non-integrability gives rise to a logarithmic correction such that 
finally \cite{Henk84a,Vojt96,Cham98,Sach99,Bran00,Oliv06,Bien12,Bien13}
\BEQ \label{gl:gtg}
\frac{g-g_c}{g_c} \sim t_g \simeq \left\{ 
\begin{array}{ll} A_< \: \sigma^{(d-1)/2}  & \mbox{\rm ~;~~ if $d<3$} \\
                  A_3 \: \sigma \ln \sigma & \mbox{\rm ~;~~ if $d=3$} \\
		          A_> \: \sigma            & \mbox{\rm ~;~~ if $d>3$}
\end{array} \right.
\EEQ
as $\sigma\to 0$ and with known constant amplitudes $A_<, A_3, A_>$, see appendix~C. 

Explicitly, the critical coupling $g_c(1,d)$ can be expressed as an integral
\BEQ \label{2.18}
g_c(1,d) = 2\pi \left( \int_0^{\infty} \!\D u\: e^{-d u^2} I_0(u^2)^d\right)^{-2}\, \frac{J}{\hbar^2} 
\EEQ
The asymptotic behaviour of $I_0(\rho)$ for $\rho$ large tells us that $g_c(1,d)>0$ is finite for $d>1$ but that $g_c(1,d)=0$ for $d\leq 1$. 
For $d=2$, the identities \cite[eq. (2.15.20.5)]{Prudnikov2}, \cite[eqs. (8.1.2),(15.1.26)]{Abra65} give the closed expression
\BEQ \label{2.19}
g_c(1,2) = \frac{16}{\pi^2} \left[ \Gamma\left(\frac{5}{8}\right)\Gamma\left(\frac{7}{8}\right)\right]^4 \, \frac{J}{\hbar^2} 
\simeq 9.67826 \, \frac{J}{\hbar^2}
\EEQ
which agrees with the numerical values quoted in \cite{Cham98,Oliv06}. The result (\ref{2.19}) 
is the counterpart to the exact value of $T_c$ in the $3D$ classical spherical model \cite{Cara03}. 

{\bf 3.} In the general case $0<\lambda<1$, the asymptotic analysis of the spherical constraint is more involved than in the two previous cases. 
As far as the critical exponents are concerned, we show in appendix~C that (\ref{gl:gtg}) remains valid for $0<\lambda\leq 1$, where the
amplitudes are given explicitly by eqs.~(\ref{C13},\ref{C14},\ref{C15}).\footnote{For $\lambda>1$, 
the asymptotic methods used in appendix~C for analysing (\ref{2.11}) cannot be taken over, since the argument
$\varrho$, see eq.~(\ref{2.12}), of the Bessel functions can vanish. 
The contributions of such zeroes would have to be included into the analysis. However, since the numerical
values do not show evidence for a singularity at $\lambda=1$, we expect that our results should be straightforwardly generalisable to $\lambda>1$.} 

\begin{figure}[tb]
\centerline{\psfig{figure=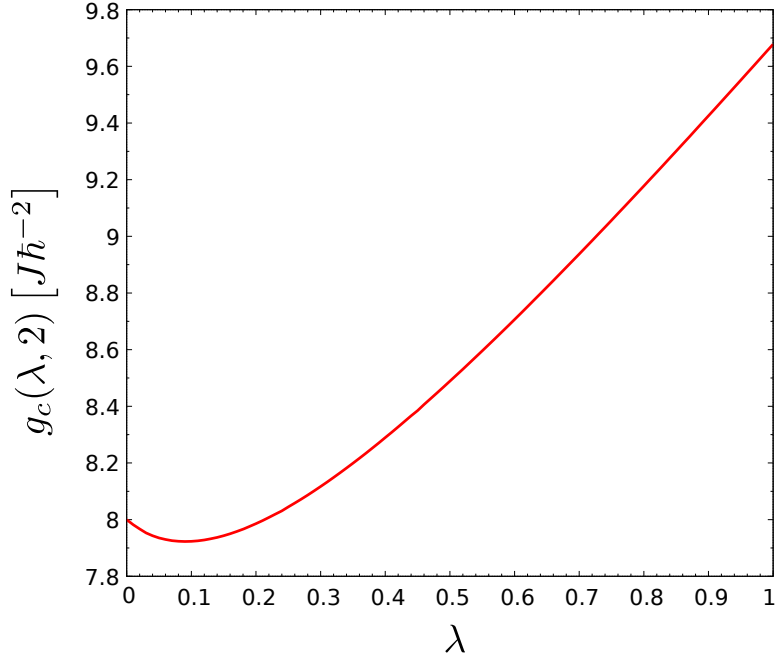,width=4.0in,clip=}}
\caption[fig1]{Critical coupling $g_c(\lambda,2)$ from (\ref{2.20},\ref{2.21}), 
as a function of the pair creation/annihilation rate $\lambda$, for $d=2$ space dimensions. \label{fig1}
}
\end{figure}

Turning to the values of the critical coupling $g_c=g_c(\lambda,d)$, we consider first the $2D$ case and have
\BEQ \label{2.20}
g(\lambda,2) = 8\pi^2 \frac{(1+\lambda)^3}{G(\lambda)^2}\, \frac{J}{\hbar^2}
\EEQ
where, using \cite[eq. (2.15.20.5)]{Prudnikov2}, we find from (\ref{2.11})
\BEA
G(\lambda) &=&\int_0^{\infty} \!\D u\, \int_0^1 \frac{\D x \exp\left(-us_c\right)}{\sqrt{x(1-x)\,}} \left[ 
\frac{1+\lambda}{2}(1+3\lambda) I_0(\varrho)^2 -\frac{1-\lambda^2}{2} I_1(\varrho)^2 +\frac{1-\lambda^2}{2\varrho} I_1(\varrho)I_0(\varrho) \right]
\nonumber \\
&=& \int_0^1 \frac{\D x}{\sqrt{x(1-x)\,}} 
\left[ \frac{1+3\lambda}{2}\: {}_2F_1\left(\demi,\demi;1;\left(1-x\frac{2\lambda}{1+\lambda}\right)^2\right) \right.
\nonumber \\
& & \left.
-\frac{1-\lambda}{16}\left(1-x\frac{2\lambda}{1+\lambda}\right)^2\: 
{}_2F_1\left(\frac{3}{2},\frac{3}{2};3;\left(1-x\frac{2\lambda}{1+\lambda}\right)^2\right) \right.
\nonumber \\
& & \left.
+ \frac{1-\lambda}{4}\: {}_3F_2\left(\demi,1,\frac{3}{2};2,2;\left(1-x\frac{2\lambda}{1+\lambda}\right)^2\right) \right]
\label{2.21}
\EEA
This quite explicit form is more easily treated numerically than the full double integral (\ref{2.11}), 
to be considered in generic dimensions $d$. 
In figure~\ref{fig1}, we plot $g_c=g_c(\lambda,2)$ over against $\lambda$. 
While the two known values (\ref{2.14},\ref{2.19}) for $\lambda=0$ and $\lambda=1$ are certainly reproduced, 
we also observe that the behaviour of $g_c(\lambda,2)$ is {\em not} monotonous in $\lambda$, but rather has a minimum around $\lambda\approx 0.1$. 
This surprising feature of a re-entrant quantum phase transition does not have an analogue in the $3D$ classical spherical model.  

\begin{figure}[tb]
\centerline{\psfig{figure=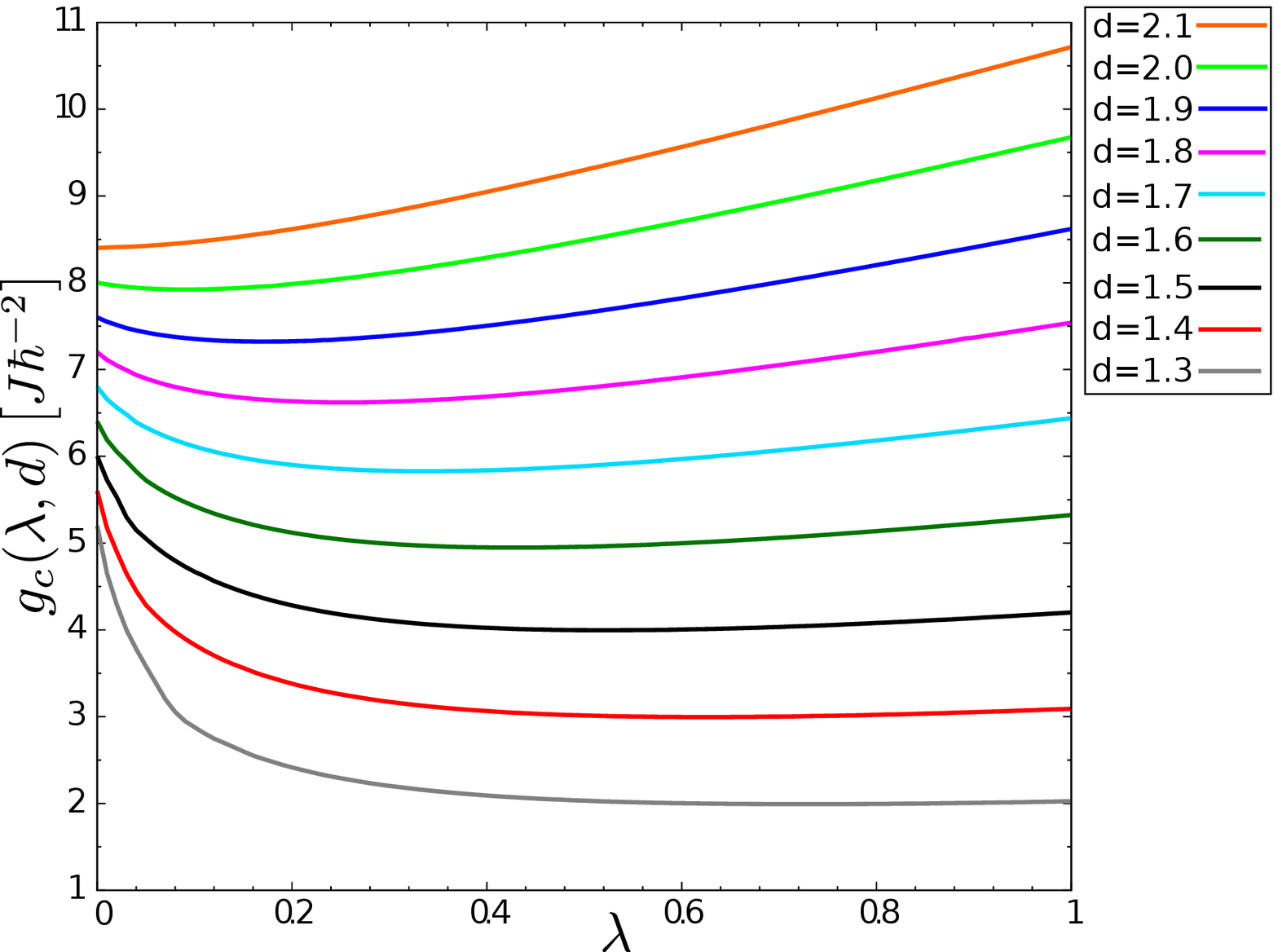,width=3.5in,clip=} ~~\psfig{figure=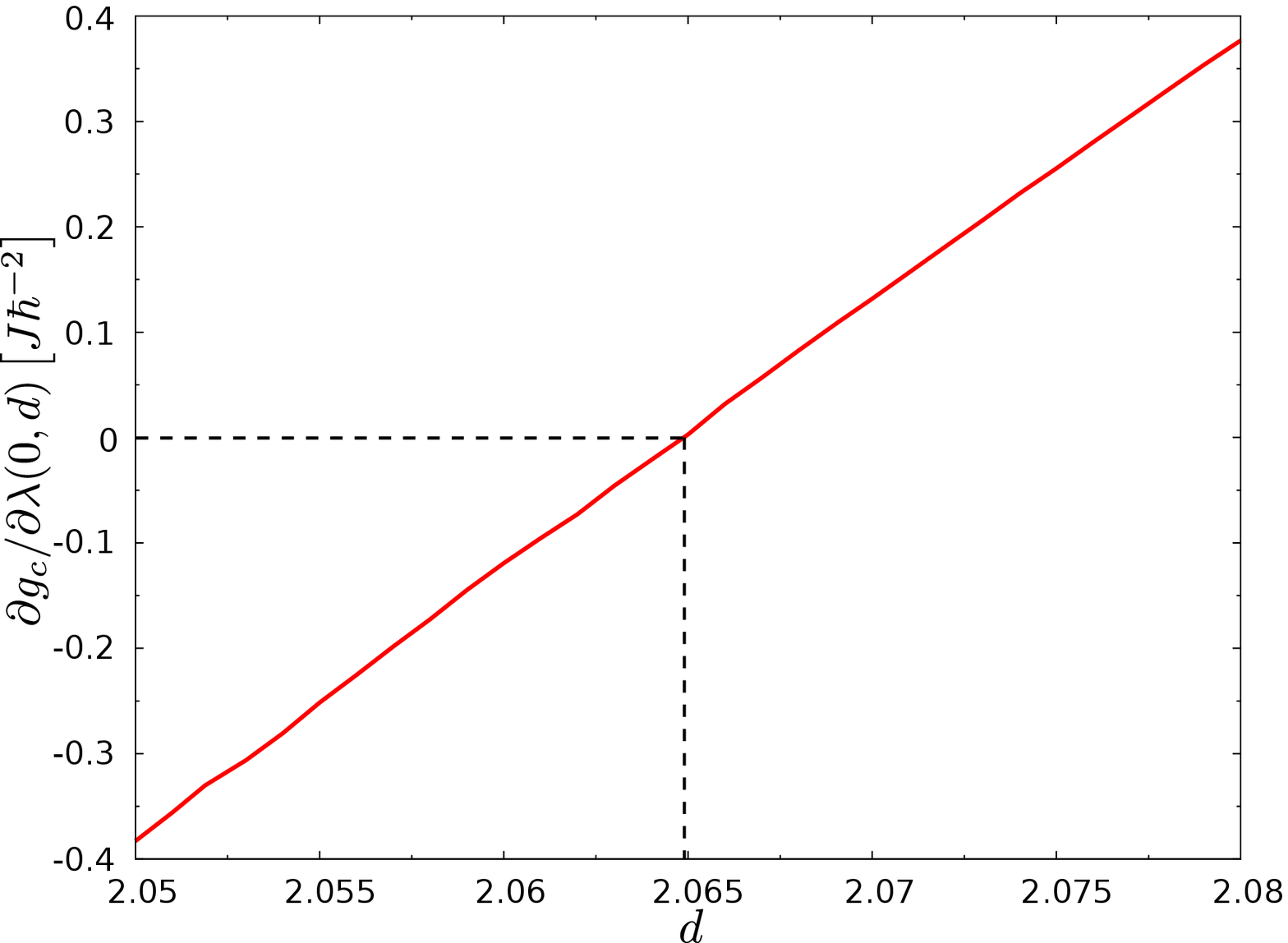,width=3.5in,clip=}}
\caption[fig2]{Left panel: Critical coupling $g_c(\lambda,d)$, computed from (\ref{2.22}), as a function of $\lambda$ for 
$d=[1.3, 1.4, 1.5, 1.6, 1.7, 1.8, 1.9, 2.0, 2.1]$ from bottom to top. 
Right panel: slope $\left.\partial g_c(\lambda,d)/\partial \lambda\right|_{\lambda=0}$ of the critical coupling
$g_c$ at $\lambda=0$, as a function of $d$. For $d\approx 2.065$, the slope vanishes. \label{fig2}
}
\end{figure}

Indeed, this re-entrant transition for $\lambda$ small enough is a generic feature of the quantum spherical model. 
In the left panel of figure~\ref{fig2}, we show the critical coupling $g_c(\lambda,d)$, as given by 
\BEA
g_c(\lambda,d) &=&  d^3 \pi^2 (1+\lambda)^3 
\left\{ \int_0^{\infty} \!\!\D u \int_0^1 \!\D x\: \frac{\exp (-u(1+\lambda)d/2)}{\sqrt{x(1-x)}\,}\, 
I_0(\varrho)^d \,\right. \nonumber \\
& & \left. \times \left[ s^2 -d(d-1)\frac{1-\lambda^2}{4} \frac{I_1(\varrho)^2}{I_0(\varrho)^2} 
- \frac{d}{2}\frac{1-\lambda^2}{4} \left( 1 + \frac{I_2(\varrho)}{I_0(\varrho)} \right) \right] \right\}^{-2} \frac{J}{\hbar^2}
\label{2.22}
\EEA
Clearly, the figure suggests that $g_c(\lambda,d)$ should go through a non-vanishing minimum for all dimensions $d\lesssim 2.1$. 

Let us make this statement more precise. First, we observe from (\ref{2.19}) that $g_c(1,2)>g_c(0,2)=8 J/\hbar^2$. Second, from (\ref{2.18}) 
it follows that $g_c(1,d)$ grows monotonously with $d$. Since $g_c(\lambda,2)$ is increasing with $\lambda$ for $\lambda$ large enough, see
figure~\ref{fig1}, this means that the slope  of $g_c(\lambda,d)$ at $\lambda=1$ should be positive, 
viz. $\left.\partial g_c(\lambda,d)/\partial \lambda\right|_{\lambda=1}>0$. 
On the other hand, in appendix~E we show that close to $\lambda=0$ one has
\BEQ \label{2.24}
g_c(\lambda,d) \simeq g_c(0,d) - \left\{
\begin{array}{ll} g_{(0)} \lambda^{d/2} & \mbox{\rm ~;~~ if $1<d<2$} \\
                  g_{(1)} \lambda       & \mbox{\rm ~;~~ if $d>2$} 
\end{array} \right. 
\EEQ
and where the known constant $g_{(0)}>0$, but the sign of the known constant $g_{(1)}$ may depend on $d$. Therefore, the slope 
$\left.\partial g_c(\lambda,d)/\partial \lambda\right|_{\lambda=0}<0$ for dimensions $1<d<2$ and diverges as $\lambda\to 0$. 
On the other hand, in  the right panel of figure~\ref{fig2}, we show the finite slope 
$\left.\partial g_c(\lambda,d)/\partial \lambda\right|_{\lambda=0}$ of $g_c$ at $\lambda=0$, for dimensions $d>2$, 
as a function of $d$. Clearly, the slope of $g_c$ at $\lambda=0$ is negative for
$d\lesssim 2.065$ and becomes positive for larger values of $d$. 
For $d$ small enough, the slope of $g_c(\lambda,d)$ is negative at $\lambda=0$ and 
positive at $\lambda=1$. By Rolle's theorem, the critical coupling $g_c(\lambda,d)$ should have
a minimum at some non-vanishing value of $\lambda$, for all dimensions $d\lesssim 2.065$. This is indeed what we observe in the left panel of
figure~\ref{fig2}. In  consequence, {\em the spin-anisotropic quantum spherical model has a re-entrant quantum phase transition for dimensions
$d\lesssim 2.065$.}

\subsection{Physical observables near quantum criticality}

The scaling of the thermodynamic observables follows from the free-energy density. Since we restrict ourselves to an 
analysis of the zero-temperature properties of our model, the quantum coupling $g$ takes over the role of the temperature
in classical spin systems, such that $t_g$ as defined in (\ref{2.15},\ref{gl:gtg}) 
takes over the role of $T-T_c$ in classical phase transitions.
Therefore, one expects for the 
singular part $f^{\rm sin}$ of the free energy density 
\BEQ \label{2.25}
f=\frac{F}{{\cal{N}}}=-\frac{B^2}{4J}\frac{1}{\sigma} +\hbar\sqrt{\frac{gJ}{2s}}\int_{\cal{B}}
\frac{\D\vec{k}}{(2\pi)^d}
\sqrt{ s - \frac{1+\lambda}{2} \sum_{j=1}^d \cos k_j\,}\:
\sqrt{ s - \frac{1-\lambda}{2} \sum_{j=1}^d \cos k_j\,}
\EEQ
to obey the following scaling behaviour 
\BEQ \label{gl:scaling_f}
f^{\rm sin}(t_g,B) = A_1 |t_g|^{2-\alpha} W_{\pm}\left( A_2  B |t_g|^{-\beta-\gamma}\right)
\EEQ
where $W_{\pm}$ are universal scaling functions, associated with the sign of $t_g\gtrless 0$, and $\alpha,\beta,\gamma$ are the standard critical
exponents. All non-universal information on the specific model can be absorbed into the two metric factors $A_{1,2}$. 
Similarly, we consider the spin-spin correlation (\ref{2.6}) $C(|\vec{r}|)=\langle S_{\vec{n}}S_{\vec{n}+\vec{r}}\rangle$ 
at zero temperature $T=0$. As shown in appendix~D, we can use spatial translation-
and rotation-invariance, and have for $\lambda>0$
\BEA 
C(R) &=& \left< S_0 S_R\right> \:=\:
\sqrt{\frac{\hbar^2g}{Js}}\int_{\cal{B}}\frac{\D \vec{k}}{(2\pi)^d}
\sqrt{\frac{2s-(1-\lambda)\sum_{j=1}^d \cos k_j}{2s-(1+\lambda)\sum_{j=1}^d \cos k_j}\,}\,
\cos k_1 R
\nonumber \\
&=&
\sqrt{\frac{\hbar^2g}{Js}} \frac{1}{(2\pi)^{(d+1)/2}}\frac{s-\frac{1-\lambda}{2}d}{\sqrt{\lambda (1+\lambda)d/2}\,}
\: \left(\frac{1}{\xi R}\right)^{(d-1)/2} K_{\frac{d-1}{2}} \left(\frac{R}{\xi}\right)
\label{correlator}
\EEA
where we identify the correlation length, with $s=\demi(1+\lambda)d+\sigma$, as follows 
\BEQ \label{gl:xicorr}
\xi =  \sqrt{\frac{1+\lambda}{4}\,}\: \sigma^{-1/2}
\EEQ
and $K_{\nu}(x)$ is the other modified Bessel function \cite{Abra65}. 
For isotropic classical phase transitions, a long-standing result of Privman and Fisher \cite{Priv84} states that there exist
only two independent non-universal metric factors, such as $A_{1,2}$. For quantum systems, anisotropies are possible between correlators
along the spatial lattice and correlations in the (euclidean) `time' direction and generated via the transfer matrix
${\cal T}=\exp\left(-\tau H\right)$. One then must distinguish `parallel' distances $r_{\|}$ along the `time' direction and
`perpendicular' distances $\vec{r}_{\perp}$ along the space direction. The correlation length $\xi=\xi_{\perp}$ considered here
is spatial, whereas the `temporal' correlation length $\xi_{\|}\sim (\Delta E)^{-1}$ is related to the energy gap of $H$. 
The anisotropy between `time' and `space' introduces a further metric factor which in those cases where there is a classical analogue, 
and therefore the dynamical exponent $z=1$, amounts simply to a further independent amplitude $D_0$ related to the freedom
of normalisation of the quantum hamiltonian $H$. For such anisotropic or quantum systems (at $T=0$), one expects a scaling form for a two-point correlator 
\cite{Henk01,Camp14,Kirk15}
\BEQ \label{gl:scaling_C}
C(R;t_g,B) = D_0 D_1 R^{2-d-z-\eta}X_{\pm}\left(|\vec{R}|/\xi; D_0r_{\parallel}/\xi^z ; D_2 B \left| t_g\right|^{-\beta-\gamma}\right)
\EEQ
where in the situation under study here, we have $R=|\vec{R}|=|\vec{r}_{\perp}|$ and $r_{\|}=0$. As before, $X_{\pm}$ are universal scaling functions 
with non-universal metric factors $D_{0,1,2}$. For isotropic systems, one has $z=1$ such that the distinction between the scaling of
$\vec{r}_{\perp}$ and $r_{\|}$ is no longer necessary and $D_0=1$ without restriction to the generality. Then, in that situation, 
only two of the four metric factors $A_{1,2}, D_{1,2}$ are independent, according to the long-standing Privman-Fisher hypothesis \cite{Priv84}.
This follows by tracing the metric factors as they occur in 
the thermodynamic observables and using the static fluctuation-dissipation theorem. 
For potentially anisotropic or quantum systems, even if $z=1$, this argument has to be generalised in order to admit a potentially
non-universal normalisation $D_0$. This leads to the following universal amplitude combinations $Q_{1,2,3}$ \cite{Henk01}
\BEA
Q_1 = A_1\xi_0^{d+z}D_0^{-1}; \ Q_2= D_2A_2^{-1}; \ Q_3=D_0^{\gamma/(\nu(d+z))}D_1A_1^{-1-\gamma/(\nu(d+z))}A_2^{-2}
\EEA
where the amplitude $\xi_0$ is from $\xi \simeq \xi_0 t_g^{-\nu}$. 
Here, we shall use the dependence on the parameter $\lambda>0$ to control explicitly the universality and hence to test the scaling forms
(\ref{gl:scaling_f},\ref{gl:scaling_C}). 

Returning to the quantum spherical model at $T=0$, the analysis of the spherical constraint, see appendix~C, 
has given us the dependence of the shift $t_g$ on the
shifted spherical parameter $\sigma=s-s_c$. Including now the magnetic field $B$ as well, we have to leading order in $\sigma$
\BEQ \label{gl:tg}
t_g - \sqrt{\frac{8J}{\hbar^2g_c}}\left(\frac{B}{2J}\right)^2 \sigma^{-2} \simeq 
\left\{ \begin{array}{ll} A_<\: \sigma^{\frac{d-1}{2}}  & \mbox{\rm   ~;~~ if $d<3$} \\ 
A_3\ \sigma \ln \sigma &\mbox{\rm    ~;~~ if  $d=3$}\\ A_>\: \sigma & \mbox{\rm   ~;~~ if $d>3$}
\end{array} \right.
\EEQ
with explicitly known amplitudes $A_<$, $A_3$ and $A_>$. 
For a non-vanishing magnetic field $B\neq 0$ the magnetic contribution will always dominate the behaviour of the
spherical constraint near criticality.

{\bf 1.} First, we treat the case $0<\lambda<1$ and $1<d<3$.
From the Gibbs free energy, eq. (\ref{2.25}), we find for the magnetisation near criticality
\BEQ
m(t_g,B) =-\frac{\partial f(t_g,B)}{\partial B}  = \frac{B}{2J}\,\frac{1}{\sigma} 
\EEQ
where the spherical constraint(\ref{gl:tg}) must be used. 
The critical behaviour is extracted by moving along the quantum critical `isochore' $B=0$ or else the quantum critical `isotherm' $t_g = 0$. 
We obtain
\BEQ
m(t_g,0) \simeq  \left[ \frac{\hbar^2g_c}{8J} \right]^{1/4} \cdot {t_g}^{1/2} \;\; , \;\;
m(0,B) \simeq A_<^{\frac{2}{d+3}}(2J)^{\frac{1-d}{d+3}}\left( \frac{\hbar^2g_c}{8J}\right)^{\frac{1}{d+3}} \cdot B^{\frac{d-1}{d+3}}
\EEQ
where we used the non-universal amplitudes from (\ref{gl:tg}) and the value of $g_c=g_c(\lambda,d)$, which are explicitly $\lambda$-dependent. 

The analogue of the susceptibility is defined by $\chi(t_g,B) = \partial m(t_g,B)/\partial B$. Explicitly, we find 
\BEA
\chi(t_g,0) &=& \frac{A_<^{\frac{2}{d-1}}}{2J}\cdot t_g^{-\frac{2}{d-1}}\\
\chi(0, B)  &=& \frac{d-1}{d+3}A_<^{\frac{2}{d+3}}(2J)^{\frac{1-d}{d+3}}\left(\frac{8J}{\hbar^2g_c}\right)^{-\frac{1}{d+3}} \cdot B^{-\frac{4}{d+3}}
\EEA

In general, the specific heat is given by the second derive of the free energy with respect to the temperature (here replaced by $t_g$). 
Here, we consider its analogue, where the role of $T$ is taken over by $t_g$. 
Furthermore, in the spherical model, the spherical constraint requires a little more careful consideration, 
which amounts to 
\BEQ
c(t_g,B)= -\frac{\partial}{\partial t_g} \left(\left.\frac{\partial f^{sin}(t_g,B)}{\partial t_g} \right|_{s}\right) \\
\EEQ
where the first derivative must be taken grand-canonically, with fixed spherical parameter, whereas the second derivative is an usual
thermodynamic derivative, in the canonical ensemble, see e.g. \cite{Berl52,Lewi52,Baxt82,Henk84a,Bran00,Bien13}. We find
\BEA
c(t_g,0)&=&  c_0+\frac{2}{d-1}\sqrt{\frac{\hbar^2g_c}{8J}} J
A_<^{-\frac{2}{d-1}} \cdot t_g^{-\frac{d-3}{d-1}}\\
c(0, B) &=& c_0 + \frac{1}{d-1}\left(\frac{\hbar^2 g_c}{8J}\right)^{\frac{1}{d+3}}\left(8J^3\right)^{\frac{d-1}{d+3}}
A_<^{-\frac{2d}{d+3}} \cdot B^{2\frac{d-3}{d+3}}
\EEA
where $c_0$ is an unimportant background constant. 

The correlation length $\xi$, introduced in eq. (\ref{gl:xicorr}), reads near criticality
\BEA
\xi(t_g,0) &=& \sqrt{\frac{1+\lambda}{4}}\, A_<^{\frac{1}{d-1}} \cdot t_g^{1/(d-1)} \\
\xi(0,B)   &=& \sqrt{\frac{1+\lambda}{4}}\, \left(\sqrt{\frac{\hbar^2g_c}{8J}}A_<\right)^{\frac{1}{d+3}}(2J)^{\frac{2}{d+3}} \cdot B^{-\frac{2}{d+3}}
\EEA
Here, the correlation length $\xi\sim 1/\Delta E$ is related to the lowest energy gap in the hamiltonian $H$, such that the dynamical exponent $z=1$. 

Finally, for the correlation function, we have from (\ref{correlator}) that at criticality, where $\sigma=0$ 
\BEQ
C(R) = \left\langle S_0 S_R\right\rangle =  
\sqrt{\frac{\hbar^2g_c}{J}} \frac{\sqrt{\lambda/2}}{1+\lambda}\,  \pi^{-\frac{1+d}{2}}\Gamma\left(\frac{d-1}{2}\right)R^{1-d}
\EEQ
In contrast to the thermodynamics observables considered before, this result\footnote{Observe that the exponents of $R$ in 
$C(R)\sim R^{-(d-1)}$ for $\xi\gg R$ and $C(R)\sim R^{-d/2} e^{-R/\xi}$ for $\xi\ll R$ are different. 
For $d=2$, one recovers the Ornstein-Zernicke form.}  
holds true for arbitrary dimensions and is not restricted to $d<3$. 

\begin{table}[tb]
\caption[tab1]{Critical exponents for the quantum spherical model (\ref{gl:Hsm2}) 
at zero temperature, along the quantum critical isochore $B=0$, in dependence on the dimension $d$ and the coupling $\lambda$. 
\label{tab1}} 
\begin{center}
\begin{tabular}{||lr|cccccc||} \hline\hline
\multicolumn{2}{||l|}{critical isochore}
                          & ~$\alpha$~    & ~$\beta$~ & ~$\gamma$~ & ~$\nu$~   & ~$\eta$~ & ~$z$~ \\ \hline\hline
$d<3$ & $\lambda\ne 0$    & $(d-3)/(d-1)$ &  $1/2$    &  $2/(d-1)$ & $1/(d-1)$ & 0        & $1$  \\ 
$d>3$ & $\lambda\ne 0$    & $0$           &  $1/2$    &  $1$       & $1/2$     & 0        & $1$  \\ \hline
$d>0$ & $\lambda= 0$      & $0$           &  $1/2$    &  $1$       & $--$      & $--$     & $2$  \\ \hline\hline
\end{tabular} \end{center}
\end{table}

For the interpretation of these results, we recall the conventional critical exponents and also the associated amplitudes, 
in the notation of \cite{Priv91},\footnote{In order to avoid ambiguities, we write $D$ for the amplitude denoted as $B$ in \cite{Priv91}, 
since we have already used the letter $B$ to denote the magnetic field. 
Analogously, along the quantum critical `isotherm' $t_g=0$, we write $D_c$ instead of the conventional notation $B_c$ \cite{Priv91}.} 
along the quantum critical `isochore' $B=0$ 
\BEA 
m \simeq D t_g^{\beta}\ ; \ 
\chi \simeq \Gamma t_g^{-\gamma}\ ; \  
c \simeq \frac{A}{\alpha} t_g^{-\alpha} + c_0 \ ; \ 
\xi \simeq\xi_0 t_g^{-\nu} \ ; \ 
G(R) \sim R^{2-d-z-\eta}\ ; \ 
\Delta E \sim \xi^{-z}
\EEA
The values of the exponents can be read off and are collected in table~\ref{tab1}. As expected they agree with those of the classical
spherical model in $d+1$ dimensions. 

Along the quantum critical isotherm, $t_g =0$, one can define 
\BEA
c \simeq c_0+ (A_c/\alpha_c) |B|^{-\alpha_c} \ ; \ \chi \simeq \Gamma_c|B|^{-\gamma_c}\ ; \ B \simeq D_c m |m|^{\delta-1} \ ; \ \xi \simeq \xi_c |B|^{-\nu_c}
\EEA
and read off the exponents,\footnote{These obey the standard scaling relations, such as $\alpha_c = {\alpha}/{\beta\delta}$, $\gamma_c = 1-1/\delta$, 
$\nu_c = {\nu}/{\beta\delta}$.} collected in table~\ref{tab2}.  
The universality of this quantum phase transition is confirmed through the $\lambda$-independence of all these exponents. 

\begin{table}[tb]
\caption[tab2]{Critical exponents for the quantum spherical model (\ref{gl:Hsm2}) 
at zero temperature, along the quantum critical `isotherm' $t_g=0$, in dependence on the dimension $d$ and the coupling $\lambda$. 
\label{tab2}} 
\begin{center}
\begin{tabular}{||lr|cccc||} \hline\hline
\multicolumn{2}{||l|}{critical isotherm}
                          & ~$\alpha_c$~     & ~$\gamma_c$~ & ~$\delta$~    & ~$\nu_c$~    \\ \hline\hline
$d<3$ & $\lambda\ne 0$    & $2(d-3)/(d+3)$   & $4/(d+3)$    & $(d+3)/(d-1)$ & $2/(d+3)$    \\
$d>3$ & $\lambda\ne 0$    & $0$              & $2/3$        & $3$           & $1/3$        \\ \hline
$d>0$ & $\lambda= 0$      & $0$              & $-1/3$~~     & $3$           & --           \\ \hline\hline			  
\end{tabular} \end{center}
\end{table}

In addition, the universality of full scaling scaling forms (\ref{gl:scaling_f},\ref{gl:scaling_C}) can be tested by working out at least three universal
amplitude combinations \cite{Priv91}. Considering the singular free energy and its derivatives, we considered three amplitude combinations which 
from (\ref{gl:scaling_f}) are expected to be universal. Explicitly
\BEA
R_c =A\Gamma/D^2&=& \frac{3-d}{(d-1)^2} \nonumber\\
R_\chi = \Gamma D_c B^{\delta-1}&=& 1 \label{2.44} \\
\delta \Gamma_c D_c^{1/\delta} &=& 1 \nonumber
\EEA
and we give the results which follow from our explicit calculations above. The $\lambda$-independence of these three amplitude ratios is additional
confirmation of the scaling form (\ref{gl:scaling_f}), with only two non-universal metric factors. 
In order to test the universality of the scaling form (\ref{gl:scaling_C}) of the spin-spin correlator, consider 
\BEA
Q_1 &=& 2^{2-d} \frac{\Gamma\left( \frac{1-d}{2}\right)}{ \Gamma\left( \frac{d-1}{2}\right)} \frac{W_+''(0)^2}{W_+'(0)^2}X_+(0)
\nonumber \\
Q_3 &=& 2^{\frac{2d}{d+1}}\left( \frac{\Gamma\left( \frac{d-1}{2}\right)}{ \Gamma\left( \frac{1-d}{2}\right)X_+(0)} \right)^{\frac{2}{d+1}}\frac{W_+'(0)^{\frac{1-d}{d+1}}}{W_+''(0)^{\frac{4}{d+1}}}
\label{2.45}
\EEA
whose universality is confirmed explicitly through the $\lambda$-independence. Observe that for $1<d<3$ all universal 
amplitude ratios in (\ref{2.44},\ref{2.45}) are finite, but that several of them they either vanish or explode when $d\to 1$ or $d\to 3$. This
indicates that the scaling behaviour is going to be different (or does not even exist) when $d\geq 3$ or $d\leq 1$. 

For the spin-anisotropic quantum spherical model, we can conclude that {\em the scaling forms 
(\ref{gl:scaling_f},\ref{gl:scaling_C}), and their universality, 
have been fully confirmed at the quantum critical point at $T=0$, $g=g_c(\lambda,d)$, with $1<d<3$ and $0<\lambda\leq 1$}. 
Since the scaling functions themselves are universal, they were already calculated explicitly in the classical spherical model
in $d+1$ dimensions, see e.g. \cite{Bran00}, and need not be repeated here.

{\bf 2.} For $0<\lambda <1$ and $d=3$, we are working at the upper critical dimension. Therefore, we have to
introduce logarithmic corrections to the scaling behaviour, see eq. (\ref{gl:tg}).
In order to work with the logarithmic terms and the magnetic field, we introduce the dimensionless field
$\wht{B}:=\sqrt{\frac{8J}{\hbar^2g_c}} \frac{B}{2J} $. In this manner, the expression $\ln \wht{B}$ is well-defined.
We find for the magnetisation
\BEA
m(t_g,0) &\simeq& \left[ \frac{\hbar^2g_c}{8J} \right]^{\frac{1}{4}}\cdot {t_g}^{\frac{1}{2}}\\
m(0,\wht{B})   &\simeq& \sqrt{\frac{\hbar^2g_c}{8J}}\left(\frac{2}{3}A_3\right)^{\frac{1}{3}} \cdot |\wht{B}|^{\frac{1}{3}}|\ln|\wht{B}||^{\frac{1}{3}}
\EEA 
and for the susceptibility
\BEA
\chi(t_g,0) &\simeq& \frac{A_3}{2J}\cdot |t_g|^{-1} |\ln|t_g||\\
\chi(0,\wht{B})   &\simeq& \frac{1}{2J}\left( \frac{2A_3}{3} \right)^{\frac{1}{3}}\cdot |\wht{B}|^{-\frac{2}{3}}|\ln|\wht{B}||^{\frac{1}{3}}
\EEA
In the same manner as above, we calculate the specific heat and find
\BEA
c(t_g,0) &\simeq& \sqrt{\frac{2\hbar^2Jg_c}{A_3^2}}\cdot |\ln|t_g||^{-1}\\
c(0,\wht{B})   &\simeq& 3\sqrt{\frac{\hbar^2Jg_c}{2A_3^2}}\cdot |\ln|\wht{B}||^{-1}\
\EEA
Finally, the correlation length reads
\BEA
\xi(t_g,0) &\simeq& \sqrt{\frac{1+\lambda}{4A_3}}\cdot |t_g|^{-\frac{1}{2}}\cdot |\ln|t_g||^{\frac{1}{2}}\\
\xi(0,\wht{B})   &\simeq& \sqrt{\frac{1+\lambda}{4}} \left( \frac{2A_3}{3}\right)^{\frac{1}{6}}\cdot |\wht{B}|^{-\frac{1}{3}}|\ln|\wht{B}||^{\frac{1}{6}}
\EEA
This logarithmic behaviour can  be described in terms of logarithmic sub-scaling exponents \cite{Kenn06}
\BEA\nonumber
c &\sim& |t_g|^{-\alpha}|\ln |t_g||^{\wht{\alpha}}\ ; \
m(t_g,0) \sim |t_g|^{\beta}|\ln|t_g||^{\wht{\beta}}\ ; \
\chi \sim |t_g|^{-\gamma}|\ln|t_g||^{\wht{\gamma}}\ ; \\
\xi &\sim& |t_g|^{-\nu}|\ln|t_g||^{\wht{\nu}}\ ; \ m(0,\wht{B}) \sim \wht{B}^{1/\delta}|\ln|\wht{B}||^{\wht{\delta}}\ ; \
C(R)\sim R^{-(d-2+z+\eta)}|\ln R|^{\wht{\eta}}
\EEA
and we simply read off their (universal, since $\lambda$-independent) values 
\BEQ
\wht{\alpha} = -1 \ ; \ \wht{\beta} = 0 \ ; \ \wht{\gamma} = 1 \ ; \ \wht{\nu} = \demi \ ; \ \wht{\delta} = \frac{1}{3}\ ; \ \wht{\eta} = 0
\EEQ 
These values agree with those of the $4D$ $O(n)$-Heisenberg model in the limit $n\rightarrow\infty$  \cite{Kenn06,Henk10}.

{\bf 3.} In the case $0<\lambda <1$ and $3<d$ we expect mean-field critical behaviour.
Near criticality $0<t_g\ll1$, we find the observables in the same manner as in the previous parts, but with the 'linear' 
spherical constraint. We find the observables along the critical $B=0$ line
\BEA
m(t_g,0) &=& \left[ \frac{\hbar^2g_c}{8J}\right]^{\frac{1}{4}} \cdot {t_g}^{\frac{1}{2}} \\
\chi (t_g,0) &=& \frac{A_>}{2J} \cdot t_g^{-1}\\
c(t_g,0) &=&\frac{1}{\sqrt{2}A_>}\sqrt{\hbar^2 J g_c}\\
\xi(t_g,0) &=&\sqrt{\frac{1+\lambda}{4}A_>} \cdot t_g^{-\frac{1}{2}}
\EEA
and along the quantum critical isotherm $t_g=0$ they read
\BEA
m(0,B) &=& \left[ \frac{1}{2JA_>}\sqrt{\frac{\hbar^2 g_c}{8J}}\right]^{\frac{1}{3}} \cdot B^{\frac{1}{3}}\\
\chi(0,B) &=& \frac{1}{2J}\left[ \frac{A_>}{(2J)^2}\sqrt{\frac{8J}{\hbar^2g_c}} \right]^{-\frac{1}{3}} \cdot B^{-\frac{2}{3}}
\label{2.61} \\
c(0,B) &=& \frac{1}{\sqrt{2}A_>}\sqrt{\hbar^2 J g_c}\\
\xi(0,B) &=& \sqrt{\frac{1+\lambda}{4}}\left[\frac{(2J)^{\frac{3}{2}}\hbar\sqrt{g_c}}{2A_>}\right]^{\frac{1}{6}} \cdot B^{-\frac{1}{3}}
\EEA
Reading off the critical exponents (see tables~\ref{tab1} and~\ref{tab2}) yields the expected mean-field behaviour. 

{\bf 4.} For $\lambda =0$ and $d$ arbitrary, the free energy density reads
\BEQ
f(t_g,B) = -\frac{B^2}{4J}\frac{1}{\sigma} +\hbar\sqrt{\frac{gJ}{2}}\sqrt{s}
\EEQ
The magnetisation reads consequently
\BEA
m(t_g,0) &\simeq& \frac{1}{\sqrt{8}} \cdot {t_g}^{1/2} \\
m(0,B)&\simeq& \left( 2Jd\right)^{-1/3} \cdot B^{1/3}
\EEA
and the magnetic susceptibility becomes
\BEA
\chi(t_g,0) &\simeq& \frac{1}{2J}\frac{\sqrt{8}}{d}\cdot t_g^{-1} \\
\chi(0,B)&\simeq& \left(\frac{d}{2J}\right)^{1/3} \cdot B^{1/3}
\EEA
The specific heat is found to be constant near criticality and along the quantum critical isotherm
\BEQ
c \simeq \frac{d^{\frac{3}{2}}}{2}J
\EEQ
The critical exponents are listed in tables~\ref{tab1} and~\ref{tab2}. They are distinct from those of the modified quantum spherical models
defined in \cite{Niew95,Grac04}, where the particle number $\wht{N}$ is conserved as well. 

For the correlation function, we see a disconnected part from the zero temperature contribution. As derived in appendix~D, we have to 
take thermal contributions into account. We then find
\BEQ
C(R) = \sqrt{\frac{\hbar^2g}{8Js}} + \sqrt{\frac{\hbar^2g}{2Js}}\exp\left(-2zs\right) I_0(z)^{d-1}I_R(z)
\EEQ
with $z= \sqrt{gJ\hbar^2/2T^2s}$. At criticality, we can deduce to leading order in $T$, see eq.~(\ref{D17})
\BEA
C(R) &=& \sqrt{\frac{\hbar^2 g_c}{4dJ}} \delta_{R,0} + \sqrt{\frac{\hbar^2g_c}{dJ}}\left(\frac{T^2d}{4\pi^2g_cJ\hbar^2} \right)^{d/4} 
\exp\left( - \frac{R^2T}{2}\sqrt{\frac{d}{g_cJ\hbar^2}} \right)
\nonumber \\
&=& \demi \frac{T}{J} \xi_{T}^{-2}\, \delta_{R,0} + \frac{T}{J} \xi_T^{2-d} \exp\left(-\demi\left(\frac{R}{\xi_T}\right)^2\right)
\label{2.71}
\EEA
with the thermal reference length $\xi_T^{-4} := T^2d/g_c J\hbar^2$ and 
where the critical coupling constant $g_c=g_c(0,d;T)$ has to be found from the spherical constraint in the non-vanishing zero-temperature limit. 
To leading order in $T$, this gives the condition
\BEQ
\sqrt{\frac{J}{\hbar^2g_c}} = \sqrt{\frac{1}{4d}\,}+\frac{d^{-1/2}}{(2\pi)^{d/2}}\left(\frac{d}{g_cJ\hbar^2} \right)^{d/4} T^{d/2}
\EEQ
hence $g_c \simeq 4d\left( 1- \frac{2/\sqrt{d}}{(4\pi)^{d/2}}\left(\frac{T}{J}\right)^{d/2}+\ldots\right)\frac{J}{\hbar^2}$, 
which illustrates how finite-temperature effects renormalise the value of $g_c$.  
The behaviour (\ref{2.71}) of the correlation function does not fit into the standard phenomenology, 
described by the conventional critical exponents \cite{diFr97,Henk99,Sach99,Bran00}. 

\subsection{Casimir effect in $d=1$ dimension}

Although an analysis of finite-size effects is beyond the scope of this work, we add a brief comment on the Casimir effect in the $d\searrow 1$ limit, 
that is the case of a strip geometry, of finite width $L$, and with periodic boundary conditions. 

For $1D$ quantum systems with sufficiently short-ranged interactions and a classical correspondent model such that $z=1$, 
conformal invariance is expected to hold at the quantum ciritcal point at temperature $T=0$, 
see \cite{diFr97,Henk99}. Scale-invariance alone gives for the normalised free energy density
$f/D_0 = f_0 - L^{-2} Y( C_1 t_g L^{1/\nu_{\perp}}, C_2 B L^{(\beta+\gamma/\nu_{\perp})}) + {\rm o}(L^{-2})$ where $Y$ is an universal scaling
function and $C_{1,2}$ and $D_0$ are the non-universal metric factors \cite{Priv84,Henk01}. The normalisation constant $D_0$ must be fixed
such that the dispersion (energy-momentum) relation becomes $E(k)= |k|$ for $k\to 0$, 
such that energy and momenta are measured in the same units, see \cite{Henk99}. 
Then conformal invariance relates the universal value $Y(0,0) = -\pi c/6$ to the central charge of the corresponding $2D$ conformal field-theory 
\cite{ABCN84}. For the quantum XY chain (\ref{gl:IXY}), $f/D_0$ has indeed been calculated, $Y(0,0)$ was shown to be universal and the central
charge $c=\demi$ was found \cite{Henk87}, as expected for a model in the universality class of the $2D$ classical Ising model \cite{diFr97,Henk99,Dutt10}. 

If we want to apply the same method to the quantum spherical model in $d\searrow 1$ dimensions, we have to take into account the possibility that
the critical value $s_c$ of the spherical parameter may acquire a finite-size correction. Explicit calculations have shown, however, that this
universal finite-size amplitude vanishes, for periodic boundary conditions, when $d\searrow 1$ \cite{Luck85,Bran00}. 
Hence, $f/D_0$ can be taken over from the free fermion representation
of the quantum XY chain, where the boson-fermion correspondence implies that periodic boundary condition 
in the even sector (to which the ground-state belongs) of the quantum spherical model corresponds 
to {\em anti}-periodic boundary conditions in the even sector 
of the fermionic model (\ref{gl:IXY_ferm}) \cite{Lieb61}. Hence, the ground-state energy of the periodic spherical model chain is identical to 
the ground-state energy of the quantum XY chain, with {\em anti-periodic}
boundary conditions. This is known to read \cite{Henk87,Henk99}
\BEQ
\frac{f}{D_0} = f_0(\lambda) - \frac{\pi}{6} \frac{1}{L^2} + \frac{\pi^3}{120} \left( \frac{1}{\lambda^2} - \frac{4}{3}\right) \frac{1}{L^4} 
+ {\rm O}(L^{-6})
\EEQ
where $f_0(\lambda)$ is an explicitly known, non-universal bulk contribution to the free energy density. 
We see that the finite-size amplitude $Y(0,0)=-\pi/6$ is $\lambda$-independent and therefore universal, 
as expected \cite{Priv84,Henk01}, but the higher-order finite-size corrections are non-universal. 
We find the value $c=1$ for the central charge in $d\searrow 1$ dimensions, as expected for a free boson. 

For dimensions $d>1$, the simplifications we could use here, in the $d\searrow 1$ limit, 
do no longer apply such that the computation of the Casimir effect is considerably more involved,
see \cite{Luck85,Cham98,Danc99,Bran00,Cara03,Cham06,Cham08,Camp14} and references therein. 
It would be interesting if recent attempts to formulate a conformal bootstrap for the $3D$ Ising model \cite{elSh12} 
could be brought to shed light on the interpretation of universal Casimir amplitudes.

\section{Conclusions}

We have explored the $T=0$ quantum critical behaviour of the spin-anisotropic quantum spherical model (\ref{gl:Hsm2}). 
One of our motivations was to be able to compare the effects of bosonic versus fermionic degrees of freedom, 
by using the information available from the quantum XY model 
\cite{Kats62,Bar71,Suzu71,Henk8487,Henk87,Chak96,Henk99,Kare00,Dutt10}. 
However, the quantum spherical model has the advantage that it can be analysed exactly
for arbitrary dimensions $d$, coupling $g$ and external field $B$, 
whereas the quantum XY model is only solved for $d=1$ and for a vanishing external field $B=0$. 
As to be expected, we have found a line (`quantum critical isochore') 
of quantum phase transitions and used the pair creation/annihilation rate $\lambda>0$ to test
explicitly for universality along this line. The generalised Privman-Fisher scaling form, adapted to quantum criticality \cite{Priv84,Henk01,Camp14}
allowed to test not only the universality of the exponents but also of certain universal amplitude rations and in consequence of the full
scaling forms (\ref{2.25},\ref{correlator}). It is known since a long time that the critical behaviour of the fermionic model
along the critical isochore is universal \cite{Henk8487,Henk87}; we obtained here the analogous result for the bosonic model. 
Merely the values of the exponents are different (in $1D$, the identified central charges also differ). In the
quantum spherical model, an analogous test can also be carried out along the quantum critical isotherm $t_g=0$. 

In the special case $\lambda=0$, the total particle number is conserved, leading to a different global symmetry and the critical behaviour
is different. It is also distinct from the spherical model variants \cite{Niew95,Grac04} with a global conservation of the number of quantum particles. 

In the {\em fermionic} quantum XY model, the ordered phase contains a sub-phase, for $0<g<d\sqrt{1-\lambda^2\,}$, 
with spatially oscillating correlation functions \cite{Bar71,Suzu71,Henk8487,Henk87,Kare00}.
This sub-phase is characterised by level crossings in the hamiltonian energy spectrum, between the even and odd spin sectors \cite{Hoeg85},
The transition line between oscillating
and non-oscillating correlators, at $g=d\sqrt{1-\lambda^2\,}$, is characterised by the
existence of certain N\'eel ground-states \cite{Kurm82}. 
We did not succeed to detect similar properties in the {\em bosonic} quantum spherical model. 

A surprising feature of the model studied here is the re-entrant quantum 
phase transition for dimensions $d\lesssim 2.065$ and sufficiently
small values of $\lambda$. This shape of the quantum critical line could 
not have been anticipated from previous studies of the classical spherical model. This makes it clear that interactions between the momenta
cannot always be absorbed into a change of variables.\footnote{Considering the leading finite-temperature corrections to the value of $g_c(\lambda,d)$, 
it can be shown that for $T$ sufficiently small, the value of $g_c$ is only slightly renormalised such that the re-entrant transition
also occurs for finite (and small) temperatures $T>0$.}   
\begin{figure}[tb]
\centerline{\psfig{figure=wald1_phasesIsing,width=3.0in,clip=} ~~\psfig{figure=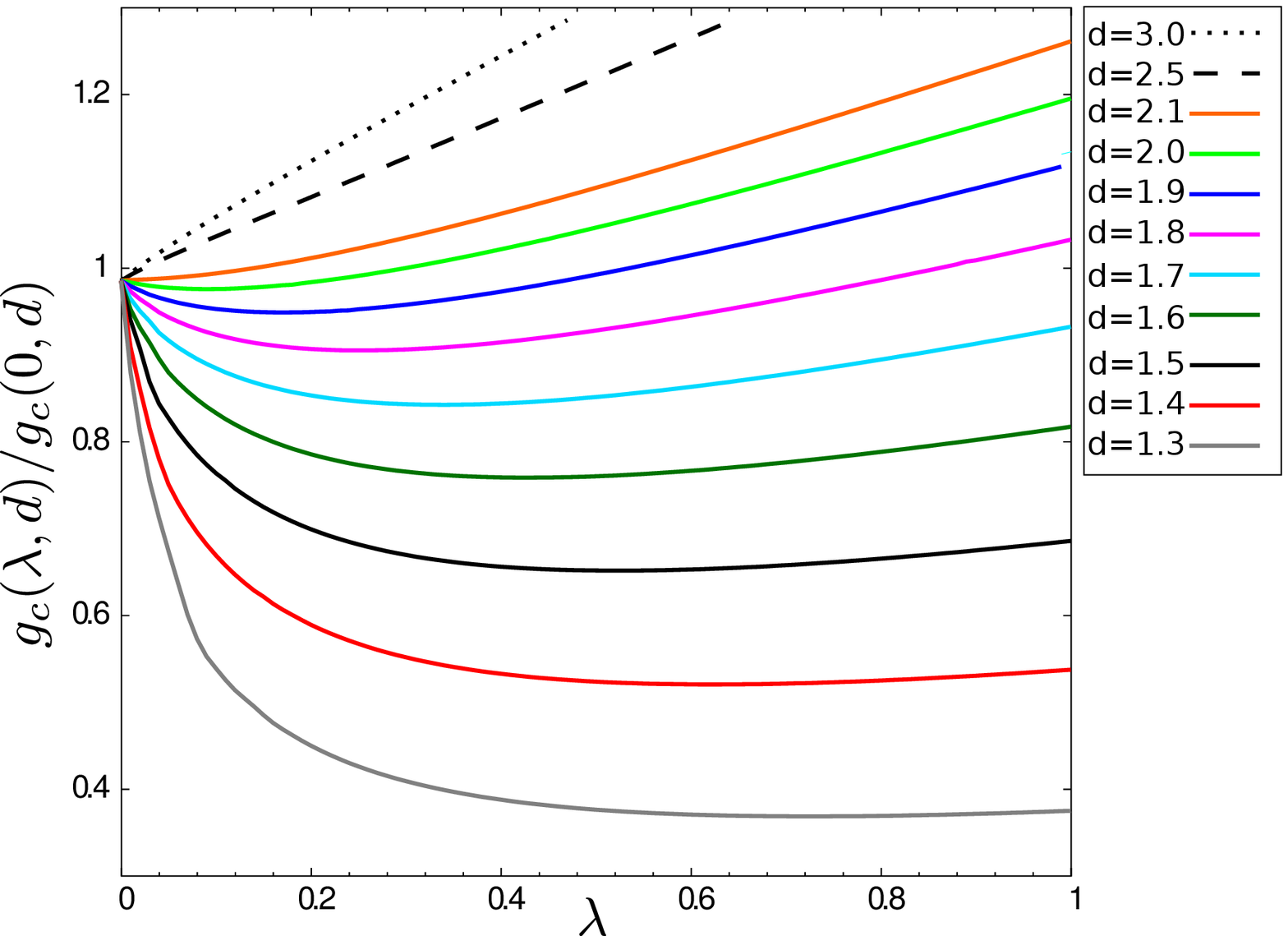 ,width=3.6in,clip=}}
\caption[fig3]{\underline{Left panel:} normalised critical coupling $g_c(\lambda)/g_c(0)$ in the quantum XY model (\ref{gl:IXY}), 
as a function of the coupling $\lambda$. In $1D$, one has $g_c(\lambda)=1$. In $2D$, 
the numerically known estimates of $g_c(\lambda)$ \cite{Henk8487} are given by the dots and the dashed line
is a guide to the eye.  
\underline{Right panel:} normalised critical coupling $g_c(\lambda,d)/g_c(0,d)$ in the quantum spherical model (\ref{gl:Hsm2}), 
as a function of $\lambda$ and for dimensions
$d=[1.3, 1.4, 1.5, 1.6, 1.7, 1.8, 1.9, 2.0, 2.1, 2.5, 3.0]$ from bottom to top. \label{fig3}
}
\end{figure}
In figure~\ref{fig3}, we compare the shape of the critical line $g_c=g_c(\lambda)$, 
normalised to the value at $g_c(0)$ at $\lambda=0$, 
of the bosonic quantum spherical model (\ref{gl:Hsm2}), 
with the fermionic quantum XY model. In $1D$, the latter model reduces to free
fermions. Comparing the shapes of $g_c(\lambda,d)$, the re-entrant 
phase transition found in the bosonic case of the {\sc saqsm} does not appear in the 
analogous $1D$ fermionic model, where $g_c(\lambda)=1$ is simply constant \cite{Bar71,Suzu71}. 
In order to better appreciate the influence of dimensionality
in the quantum XY chain on $g_c(\lambda)$, and in the absence of an analytic solution, 
the best what we can do is to compare with the
few known numerical values of $g_c(\lambda)$ in extension of the spin hamiltonian 
$H_{\rm XY}$ from (\ref{gl:IXY}) to $2D$ \cite{Henk8487}. 
Although those few data shown in figure~\ref{fig3} seem to indicate that
the approach of $g_c(\lambda)$ towards the $\lambda=0$ case should be monotonous 
and hence no re-entrant transition is suggested, the available data are too
few and too far apart for a final conclusion.   

Since in many respects, effectively non-integer values of the dimension $d$ can also 
be produced by long-ranged interactions \cite{Vojt96,Bran00,Grac04,Camp09,Flor14}, one could
anticipate that several of our conclusions might have qualitative analogues in long-ranged quantum phase transitions. 
Also, it would be interesting to see if the theory of random matrices, so sucessfully used in fermionic quantum chains \cite{Altl97,Hutc15},
could be brought to be applied to the kind of bosonic systems analysed here. 

This illustrates that the interactions between the conjugate momenta can play a physically important role. Our results raise the
question of the quantitative importance of more general kinetic terms, e.g. in O($n$)-symmetric quantum rotor models with $n$ finite. 
Also, one may anticipate a rich phenomenology when combining different kinds of interactions between the spins and the momenta. 
If such effects should be found, the spherical model would have demonstrated once more its usefulness as a heuristic device and guide
towards non-trivial and interesting new types of critical behaviour. 

\newpage
\appsection{A}{Diagonalisation via a canonical transformation} 
\label{ap:ct}
The quantum hamiltonians to be diagonalised are of the form
\BEQ
H = \sum_{n,m} \left[ \wht{a}_n^{\dag} A_{nm} \wht{a}_m 
- \demi \left( \wht{a}_n B_{nm} \wht{a}_m + \wht{a}_n^{\dag} B_{nm} \wht{a}_m^{\dag} \right) \right]
+\sum_n C_n \left( \wht{a}_n + \wht{a}_n^{\dag} \right)
\EEQ
where the sums run over the ${\cal N}=N^d$ sites of a $d$-dimensional hyper-cubic lattice. 
$A$ is a hermitian matrix, $B$ a symmetric matrix and $C$ is a real, constant vector.  
The bosonic annihilation and creation operators $\wht{a}_n, \wht{a}_n^{\dag}$ obey the standard commutators
\BEQ
{} \left[ \wht{a}_n, \wht{a}_m \right] \:=\: 
\left[ \wht{a}_n^{\dag}, \wht{a}_m^{\dag} \right] \:=\: 0 \;\; , \;\;
\left[ \wht{a}_n^{\dag}, \wht{a}_m \right] \:=\: \delta_{nm}
\EEQ
For $C=0$, the diagonalisation procedure follows
closely the fermionic techniques of Lieb, Schultz and Mattis \cite{Lieb61}, 
applied to quantum Ising/XY chains. In the bosonic case, any space 
dimension $d$ can be treated and $C\ne 0$ is admissible. 
Throughout, we restrict to the case when $A$ and $B$ are real-valued
(although extensions are readily formulated). 

We seek a canonical transformation which brings $H$ to the form
\BEQ \label{A3}
H = \sum_{k} \Lambda_k \left( \wht{b}_k^{\dag} \wht{b}_k + \demi \right) + H_0
\EEQ
where $\wht{b}_k, \wht{b}_k^{\dag}$ are again bosonic annihilation/creation operators, 
$\Lambda_k$ are the sought eigenvalues  
and the constant $H_0$ has to be determined. The required canonical transformation is of the form
\BEQ \label{A4}
\wht{b}_k = u_k + \sum_p \left( V_{kp} \wht{a}_p + W_{kp} \wht{a}_p^{\dag}\right) \;\; , \;\;
\wht{b}_k^{\dag} = u_k + \sum_p \left( V_{kp} \wht{a}_p^{\dag} + W_{kp} \wht{a}_p\right)
\EEQ
where the  ${\cal N}\times{\cal N}$ matrices ${V}$ and ${W}$ are determined from the bosonic commutation relations 
and the $u_k$ are numbers. This gives
${V}{V}^T - {W}{W}^T={\bf 1}_{d}$ and ${V}{W}^T - {W}{V}^T= {\bf 0}$, where $^T$ denotes the transpose and 
${\bf 1}_{d}$ is the $N^d \times N^d$ unit matrix. 
A direct consequence of these is $({V}+{W})({V}-{W})^T={\bf 1}_{d}$, hence
\BEQ
({V}+{W})^{-1} = ({V}-{W})^T \;\; , \;\;   ({V}-{W})^{-1} = ({V}+{W})^T
\EEQ
The last conditions on $V,W$ come from the requirement that the canonical transformation (\ref{A4}) 
brings $H$ to its diagonal form (\ref{A3}), which means $[\wht{b}_k,H]=\Lambda_k \wht{b}_k$. Hence
\BEQ \label{A6}
\left( V\pm W\right) \left( A\pm B\right) = \wht{\Lambda} \left( V \mp W\right) \;\; , \;\; 
\left( V - W\right) C = \wht{\Lambda} u
\EEQ
where $\wht{\Lambda}={\rm diag}(\Lambda_1,\ldots,\Lambda_{{\cal N}})$ is a diagonal matrix with the eigenvalues $\Lambda_k$, 
and the vector $u=(u_1\ldots,u_{{\cal N}})$. 
Following \cite{Lieb61}, one defines two matrices, arranged as two sets of vectors $(\vec{\Phi}_k)_m := (V+W)_{km}$ 
and $(\vec{\Psi}_k)_m := (V-W)_{km}$ so that by reading eq. (\ref{A6}) line by line, one has the two coupled equations
\BEQ
\vec{\Phi}_k^T \left( A + B\right) = \Lambda_k \vec{\Psi}_k^T \;\; , \;\; 
\vec{\Psi}_k^T \left( A - B\right) = \Lambda_k \vec{\Phi}_k^T
\EEQ
so that the eigenvalues $\Lambda_k$ can be found from the following eigenvalue equation\footnote{The only difference with respect
to fermionic chains \cite{Lieb61} is that therein $B=-B^T$ is antisymmetric. For bosonic as well as for fermionic systems, the
matrix $(A-B)(A+B)$ is symmetric and positive semi-definite, such that all eigenvalues $\Lambda_k$ are real.}
\BEQ \label{A8}
\vec{\Psi}_k^T M := \vec{\Psi}_k^T \left( A - B\right) \left( A + B\right) = \Lambda_k^2 \vec{\Psi}_k^T
\EEQ
Later on, we shall also need the explicit transformation of the creation/annihilation operators. For $C_n=0$, this reads 
$\wht{b}= V \wht{a} + W\wht{a}^{\dag}$ and its inverse becomes
$\wht{a}=V^T \wht{b} -W^{T} \wht{b}^{\dag}$, along with the hermitian conjugates. 
We shall require this below for the calculation of correlators. 

Next, we must find the eigenvalues $\Lambda_k$ for the specific hamiltonian (\ref{gl:Hsm2}) in the main text, 
with nearest-neighbour interactions\footnote{The method outlined in this appendix works for arbitrary interactions, 
although the practical calculations can become more involved.}.
{\it Then the diagonal form of the hamiltonian (\ref{gl:Hsm2}) is given by (\ref{A3}), 
where the eigenvalues $\Lambda_{\vec{k}}=s\bar{\Lambda}_{\vec{k}}$ are, for a hyper-cubic square of ${\cal N}=N^d$ sites in 
$d$ spatial dimensions and with periodic boundary conditions}
\BEQ \label{A9}
\bar{\Lambda}_{\vec{k}} = \sqrt{ 1 - \frac{1+\lambda}{2s} \sum_{j=1}^d \cos k_j} \: \sqrt{ 1 - \frac{1-\lambda}{2s} \sum_{j=1}^d \cos k_j}
\EEQ
{\it where the quasi-momenta $k_j= \frac{2\pi}{N} n_j$, with $n_j=0,1,\ldots N-1$ and $j=1,\ldots,d$} and the spherical parameter $s$.
 \\

\noindent
{\bf Proof:} Eq.~(\ref{A9}) can be derived from the properties of cyclic matrices \cite{Altro01} and using mathematical induction
over the dimension $d$. 
In what follows, we denote a cyclic ${\cal N}\times{\cal N}$ matrix, generated from a vector $(v_1,\ldots,v_{{\cal N}})$, by
\BD
\mathfrak{C}\left( v_1,\ldots, v_{\cal N}\right) := 
\left( \begin{array}{cccccc} 
v_1        & v_2 & v_3 & \cdots & v_{{\cal N}-1} & v_{\cal N}     \\
v_{\cal N} & v_1 & v_2 & \cdots & v_{{\cal N}-2} & v_{{\cal N}-1} \\
\vdots     &     &     & \ddots &                & \vdots         \\
v_2        & v_3 & v_4 & \cdots & v_{\cal N}     & v_1 
\end{array} \right)
\ED

For the sake of this proof, we work with the reduced, dimensionless hamiltonian $H_r := H/(\hbar\sqrt{g\mu})$. 

\textbf{Step 1}:
For $d=1$, ${\cal N}=N$. The matrices $A=A^{(1)}$ and $B=B^{(1)}$ are (the index refers to the value of $d$)
\BD
A^{(1)} = \mathfrak{C}\left( 1 , -\frac{1}{4s}, 0 ,\ldots , 0, -\frac{1}{4s}\right) \;\; , \;\;
B^{(1)} = \mathfrak{C}\left( 0 , \frac{\lambda}{4s}, 0 ,\ldots , 0, \frac{\lambda}{4s}\right)
\ED
and therefore
\BD
M^{(1)}= \mathfrak{C}\left( 1 + \frac{1-\lambda^2}{8s^2}, -\frac{1}{2s}, 
\frac{1-\lambda^2}{16 s^2}, 0 , \ldots, 0, \frac{1-\lambda^2}{16 s^2}, -\frac{1}{2s}\right) 
\ED
is cyclic as well \cite{Altro01}. The eigenvalue equation (\ref{A8}) can now be solved by the ansatz 
$(\Psi_k)_n= e^{\II kn}$. Since the cyclicity of all
matrices implies periodic boundary conditions, this produces (\ref{A9}) for $d=1$ 
and the values of $k$ are indicated.\footnote{For $k_j$ with $j\ne 0,N/2$, the eigenvalues $\Lambda_{{k_j}}=\Lambda_{k_{N-j}}$
are degenerate. so that the corresponding eigenvectors can always be chosen with real-valued components.}  
\\ 
\textbf{Step 2}:
In order to demonstrate the passage from $d$ to $d+1$ dimensions, consider a multi-index notation in $d+1$ dimensions
\BD
\vec{n} = (n_1,n_2,\ldots,n_d, n_{d+1}) = (\wit{\vec{n}} n_{d+1}) \;\; , \;\; \wit{\vec{n}}=(n_1,n_2,\ldots,n_d)
\ED
where individually, $n_j=0,1,\ldots,N-1$, with $j=1,2,\ldots,d,d+1$. 
In $d+1$ dimensions, the hamiltonian can be brought to a block form as follows
\BEA
H^{(d+1)} &=& \sum_{\vec{n}} \left\{ \left[ \wht{a}_{\vec{n}}^{\dag} \wht{a}_{\vec{n}} + \demi \right] 
-\frac{1}{4s} \sum_{j=1}^{d+1} 
\left[ \lambda\left( \wht{a}_{\vec{n}}\wht{a}_{\vec{n}+\vec{e}_j} 
+ \wht{a}_{\vec{n}}^{\dag}\wht{a}_{\vec{n}+\vec{e}_j}^{\dag}\right) 
+ \wht{a}_{\vec{n}}^{\dag}\wht{a}_{\vec{n}+\vec{e}_j} + \wht{a}_{\vec{n}}\wht{a}_{\vec{n}+\vec{e}_j}^{\dag} \right]\right\}
\nonumber \\
&=& \sum_{\wit{\vec{n}}} \sum_{k=0}^{N-1} \left\{ \left[ \wht{a}_{\wit{\vec{n}}k}^{\dag} a_{\wit{\vec{n}}k} + \demi \right] \right.
\nonumber \\
& & \left. -\frac{1}{4s} \sum_{j=1}^{d+1} 
\left[ \lambda\left( \wht{a}_{\wit{\vec{n}}k}\wht{a}_{\wit{\vec{n}}k+\wit{\vec{e}}_jk} 
+ \wht{a}_{\wit{\vec{n}}k}^{\dag}\wht{a}_{\wit{\vec{n}}k+\wit{\vec{e}}_jk}^{\dag}\right) 
+ \wht{a}_{\wit{\vec{n}}k}^{\dag}\wht{a}_{\wit{\vec{n}}k+\wit{\vec{e}}_jk} 
+ \wht{a}_{\wit{\vec{n}}k}\wht{a}_{\wit{\vec{n}}k+\wit{\vec{e}}_jk}^{\dag} \right]\right.
\nonumber \\
& & \left. -\frac{1}{4s} 
\left[ \lambda\left( \wht{a}_{\wit{\vec{n}}k}\wht{a}_{\wit{\vec{n}}k+1} 
+ \wht{a}_{\wit{\vec{n}}k}^{\dag}\wht{a}_{\wit{\vec{n}}k+1}^{\dag}\right) 
+ \wht{a}_{\wit{\vec{n}}k}^{\dag}\wht{a}_{\wit{\vec{n}}k+1} 
+ \wht{a}_{\wit{\vec{n}}k}\wht{a}_{\wit{\vec{n}}k+1}^{\dag} \right]\right\}
\nonumber \\
&=& \sum_{k=0}^{N-1} \left\{ H_k^{(d)} -\frac{1}{4s} \sum_{\wit{\vec{n}}}
\left[ \lambda\left( \wht{a}_{\wit{\vec{n}}k}\wht{a}_{\wit{\vec{n}}k+1} 
+ \wht{a}_{\wit{\vec{n}}k}^{\dag}\wht{a}_{\wit{\vec{n}}k+1}^{\dag}\right) 
+ \wht{a}_{\wit{\vec{n}}k}^{\dag}\wht{a}_{\wit{\vec{n}}k+1} 
+ \wht{a}_{\wit{\vec{n}}k}\wht{a}_{\wit{\vec{n}}k+1}^{\dag} \right] \right\} 
\nonumber \\
&=& \sum_{\wit{\vec{n}},\wit{\vec{m}}} \sum_{k,\ell=0}^{N-1} 
\left\{      \wht{a}_{\wit{\vec{n}}k}^{\dag} A_{\wit{\vec{n}}k,\wit{\vec{m}}\ell}^{(d+1)} \wht{a}_{\wit{\vec{m}}\ell} 
-\demi\left[ \wht{a}_{\wit{\vec{n}}k}        B_{\wit{\vec{n}}k,\wit{\vec{m}}\ell}^{(d+1)} \wht{a}_{\wit{\vec{m}}\ell}
           + \wht{a}_{\wit{\vec{n}}k}^{\dag} B_{\wit{\vec{n}}k,\wit{\vec{m}}\ell}^{(d+1)} \wht{a}_{\wit{\vec{m}}\ell}^{\dag}\right] 
\right\}
\nonumber
\EEA
where $H_k^{(d)}$ is the local hamiltonian in the $k$-th $d$-dimensional layer. The interaction matrices have the block
structure
\BEA
A_{\wit{\vec{n}}k,\wit{\vec{m}}\ell}^{(d+1)} 
&=& A_{\wit{\vec{n}},\wit{\vec{m}}}^{(d)}\delta_{k,\ell} -\frac{1}{4s}\left(\delta_{\ell,k+1} +\delta_{\ell,k-1}\right){\bf 1}_{d}
\nonumber \\
B_{\wit{\vec{n}}k,\wit{\vec{m}}\ell}^{(d+1)} 
&=& B_{\wit{\vec{n}},\wit{\vec{m}}}^{(d)}\delta_{k,\ell} 
+\frac{\lambda}{2s}\left(\delta_{\ell,k+1} +\delta_{\ell,k-1}\right){\bf 1}_{d}
\nonumber
\EEA
where ${\bf 1}_d$ is the $N^d\times N^d$ unit matrix.  In turn, they may be written as cyclic matrices of blocks
\BD
A^{(d+1)} = \mathfrak{C}\left( A^{(d)}, -\frac{1}{4s}, 0, \ldots, 0,-\frac{1}{4s} \right) \;\; , \;\;
B^{(d+1)} = \mathfrak{C}\left( B^{(d)}, \frac{\lambda}{2s}, 0, \ldots, 0,\frac{\lambda}{2s} \right)
\ED
Next, we write down the block structure of the eigenvalue equation (\ref{A8})
\BEA
M^{(d+1)}= \mathfrak{C}\left( 
M^{(d)}+\frac{1-\lambda^2}{8s}, 
-\frac{A^{(d)}+ \lambda B^{(d)}}{2s}, \frac{1-\lambda^2}{16 s^2}, 0 , \ldots, 
0, \frac{1-\lambda^2}{16 s^2}, 
-\frac{A^{(d)}+ \lambda B^{(d)}}{2s} \right)
\nonumber
\EEA 
Now, the habitual ansatz $\vec{\Psi}_{\wit{\vec{n}}\ell}=\vec{\Psi}_{\wit{\vec{n}}}\, e^{\II k \ell}$ where by induction hypothesis, 
$\vec{\Psi}_{\wit{\vec{n}}}$ is the eigenvector of the $d$-dimensional problem, gives for the eigenvalue in $d+1$ dimensions
\BEA
\left( \bar{\Lambda}_{\wit{\vec{k}}k_{d+1}}^{(d+1)} \right)^2 &=& \left(\bar{\Lambda}_{\wit{\vec{k}}}^{(d)} \right)^2
+\frac{1-\lambda^2}{4 s^2} \cos^2 k_{d+1} - \frac{\cos k_{d+1}}{s} \left[ 1 - \frac{1-\lambda^2}{2s} \sum_{j=1}^d \cos k_j\right]
\nonumber \\
&=& \left( \bar{\Lambda}_{\wit{\vec{k}}}^{(d)} \right)^2 +\frac{1-\lambda^2}{4 s^2} \cos^2 k_{d+1} 
\nonumber \\
& & - \frac{\cos k_{d+1}}{s} \left[ 
  \left(1 - \frac{1+\lambda}{2s} \sum_{j=1}^d \cos k_j\right) \frac{1-\lambda}{2}
+ \left(1 - \frac{1-\lambda}{2s} \sum_{j=1}^d \cos k_j\right) \frac{1+\lambda}{2}\right] 
\nonumber \\
&=& \left(1 - \frac{1+\lambda}{2s} \sum_{j=1}^{d+1} \cos k_j\right) \: \left(1 - \frac{1-\lambda}{2s} \sum_{j=1}^{d+1} \cos k_j\right)
\nonumber
\EEA
where in the last step the induction hypothesis (\ref{A9}) was used for $\bar{\Lambda}_{\wit{\vec{k}}}^{(d)}$ in $d$ dimensions.
This completes the proof. \hfill q.e.d.

Finally, we find the constant $H_0$ in (\ref{A3}). 
For the sake of notational simplicity, we only treat the case $d=1$ explicitly, but we shall give the generic result at the end. 
Since the eigenvalues are generically two-fold degenerate, we first go over to real-valued combinations
\BEQ
\left(\bar{\vec{\Psi}}_k\right)_n := \left\{ \begin{array}{ll}
\demi c_k \left[ (\vec{\Psi}_k)_n + (\vec{\Psi}_k^*)_n  \right] = c_k \cos nk & \mbox{\rm ~;~~ if $k<N/2$} \\
\frac{1}{2\II} c_k \left[ (\vec{\Psi}_k)_n - (\vec{\Psi}_k^*)_n \right] = c_k \sin nk & \mbox{\rm ~;~~ if $k>N/2$} 
\end{array} \right.
\EEQ
Here, $c_k$ is a constant which will provide appropriate normalisation. 

{}From (\ref{A6}), we further have $\bar{\vec{\Phi}}_k=\bar{\Lambda}_k^{-1} (A-B)\bar{\vec{\Psi}}_k$, hence
\BEQ
\left(\bar{\vec{\Phi}}_k\right)_n = c_k\bar{\Lambda}_k^{-1} \left( 1 - \frac{1+\lambda}{2s}\cos k\right) 
\left\{ \begin{array}{ll} \cos nk & \mbox{\rm ~;~~ if $k<N/2$} \\
                          \sin nk & \mbox{\rm ~;~~ if $k>N/2$}
\end{array} \right.
\EEQ
The normalisation constants follow from the bosonic commutator relations and which require
\BEQ
\sum_n V_{kn}^2 - W_{kn}^2 = \sum_n \left( \bar{\vec{\Phi}}_k\right)_n \left( \bar{\vec{\Psi}}_k\right)_n =1
\EEQ 
so that finally
\BEQ
c_k^2 = \frac{\bar{\Lambda}_k}{1-(1+\lambda)(2s)^{-1} \cos k} \left\{ 
\begin{array}{ll} 1/N & \mbox{\rm ~;~~ if  $k=0,N/2$} \\
2/N & \mbox{\rm ~;~~ else}
\end{array} \right.
\EEQ
The extension to $d>1$ dimensions is now obvious. 

While this gives the general method, we now apply it to the specific hamiltonian (\ref{gl:Hsm2}) 
in the main text. For a spatially constant magnetic field, all constants are equal $C_n=C$. {}From eq.~(\ref{A6}), we deduce
\BEQ
u_k=\frac{C}{\bar{\Lambda}_k} \sum_{n=0}^{N-1} (\vec{\Psi}_k)_n \ .
\EEQ
Using the geometric sum, it is obvious that $u_k$ vanishes for $k\neq 0$. For $k=0$, we find

\BEQ
u_0 = C N\bar{\Lambda}_0^{-1} c_0 
\EEQ
Thus, we are now able to write down the constant $H_0$ by rewriting the diagonal hamiltonian in the form
$ 
H_r = \demi \sum_{k} \bar{\Lambda}_k \left( \wht{b}_k^{\dag} \wht{b}_k + \wht{b}_k\wht{b}_k^{\dag} \right) + H_0
$, 
using the transformation formula (\ref{A4}) and comparing the constant terms. We find $H_0 = -\Lambda_0 u_0^2$ and hence
the ground-state energy reads
\BEQ \label{A16}
E_0 = -\bar{\Lambda}_0 u_0^2+\demi\sum_{k} \bar{\Lambda}_k =-\frac{C^2 N}{1-(1+\lambda)(2s)^{-1}}+\demi\sum_{k}\bar{\Lambda}_k   
\EEQ
with $k=\frac{2\pi}{N}n$ and $n=0,1,\ldots N-1$. The generalisation of (\ref{A16}) to $d>1$ is obvious.

\appsection{B}{Spherical constraint for $\lambda\ne 0,1$} 

We derive the spherical constraint eqs.~(\ref{2.11},\ref{2.12}) in the main text, for general $\lambda\ne 0,1$. 

Since the magnetic field term in (\ref{2.8}) is just additive, we can set $B=0$ for our purpose.  

Starting from the form (\ref{2.8}) of the spherical constraint, 
the product of the two square roots in the denominator is folded into a
single factor by the Feynman identity, see e.g. \cite{Amit84}
\begin{equation} \label{B1}
\frac{1}{\sqrt{A\,}} \frac{1}{\sqrt{B\,}} = \frac{1}{\pi} \int_0^1 \!\D x\, \frac{1}{\sqrt{x(1-x)}} \frac{1}{xA + (1-x)B} \ ,
\end{equation}
so that the constraint becomes (with the Brillouin zone ${\cal B}=[-\pi,\pi]^d$) 
\BEQ \label{B2}
\sqrt{\frac{8\pi^2 J}{\hbar^2\, g}} = s^{-\frac{3}{2}}
\int_0^1 \frac{\D x}{\sqrt{x(1-x)}}
\int_{\mathcal{B}} \frac{\D \vec{k}}{(2\pi)^d} \frac{s^2-\frac{1-\lambda^2}{4}\left[\sum_{j=1}^d \cos k_j \right]^2}
 {x\left( s- \frac{1+\lambda}{2} \sum_{j=1}^d \cos k_j \right)+(1-x)\left( s- \frac{1-\lambda}{2} \sum_{j=1}^d \cos k_j \right)}
\EEQ
However, we are looking for a representation which factorises in the momenta $k_j$, such that the dimension $d$ can be treated as a real 
parameter, in analogy to the known representations valid for $\lambda=1$. The denominator could be simply exponentiated, via the
identity $G^{-1} = \int_0^{\infty}\!\D u\, e^{-Gu}$, but the terms in the numerator still couple the different $k_j$. One might consider
to obtain these factors by deriving the exponential with respect to $x$ or $1-x$, but this cannot be done immediately, since the presence
of $x$ in both terms in the exponential would generate unwanted contributions. It is better to introduce first an auxiliary variable
\begin{equation}
y = 1-x 
\end{equation}
and to render it formally independent from $x$, by inserting a Delta function into an additional integration over $y$, according to  
\begin{equation}
\int_0^1 \D y \ \delta(y-1+x) f(x,y) = f(x,1-x) \;\; , \;\; \mbox{\rm for $0 < x < 1$.} 
\end{equation}
Now, changing the order of integrations, we can indeed re-write the denominator as an exponential and afterwards
express the numerator as a derivative of this exponential. This is done by defining the differential operator
\begin{equation}
\mathcal{D}_{xy} := s\left(-\frac{1}{u}\frac{\partial}{\partial x} -\frac{1}{u}\frac{\partial}{\partial y} 
-\frac{1}{su^2}\frac{\partial^2}{\partial x\partial y}  \right) \ .
\end{equation}
Then eq.~(\ref{B2}) can be re-written as follows 
\BEA
\sqrt{\frac{8\pi^2 J}{\hbar^2\,g}} &=& 
s^{-\frac{3}{2}}\int_0^1 \!\D x\int_0^1 \!\D y \int_{\mathcal{B}}\frac{\D \vec{k}}{(2\pi)^d}
\int_0^\infty \!\D u\:  \frac{\delta(y+x-1)}{\sqrt{xy\,}\,}  \times
\nonumber \\
& & \times \mathcal{D}_{xy} 
\exp\left[-ux \left( s- \frac{1+\lambda}{2} \sum_{j=1}^d \cos k_j \right)-uy \left( s- \frac{1-\lambda}{2} \sum_{j=1}^d \cos k_j \right)\right]
\nonumber \\
&=&  s^{-3/2} \int_0^1 \!\D x \int_0^1 \!\D y \int_0^\infty \!\D u\:  \frac{\delta(y+x-1)}{\sqrt{xy\,}\,}  \times \nonumber \\
& & \times \mathcal{D}_{xy} \bigg\{ 
\exp \left[ -u(x+y)s \right]I_0\left(ux\frac{1+\lambda}{2}+uy\frac{1-\lambda}{2}\right)^d \bigg\} \ ,
\label{B6}
\EEA
since now the integrations of the $k_j$ factorise and can be carried out separately. 
Here and below, the $I_n(\varrho)$ denote modified Bessel functions \cite{Abra65}. 

Here, a further comment is necessary concerning the argument of the modified Bessel function. 
Clearly, and taking into account that $y=1-x$ will have to 
be put back, the argument vanishes linearly at 
\begin{equation}
x_0 = \frac{1}{2}\left( 1-\lambda^{-1} \right) \ .
\end{equation}
For $0<\lambda<1$, one has $x_0<0$ which is outside the interval of integration and need not concern us. 
But for $\lambda\geq 1$, one would have $0\leq x_0\leq \demi$ inside the integration interval of $x$, 
since the derivatives of $I_0$ lead to higher order modified Bessel functions $I_n$ with $n\geq 1$, which vanish
for a vanishing argument. Then a more careful distinction of cases which takes these zeroes into account will become necessary. 

We now apply the operator $\mathcal{D}_{xy}$ to the integrand in (\ref{B6}) and also define 
\begin{equation} \label{B8}
\varrho := \varrho(u,x,\lambda) =u\left(x\frac{1+\lambda}{2}+(1-x)\frac{1-\lambda}{2}\right)
\end{equation}
Then the spherical constraint (\ref{B6}) becomes 
\BEA
\sqrt{\frac{8\pi^2 J}{\hbar^2\, g}} &=& s^{-3/2} 
\int_0^\infty \!\D u \int_0^1 \!\D x\: \frac{\exp\left[-us\right]}{\sqrt{x(1-x)\,}\,} \left[I_0(\varrho)\right]^d \times
\nonumber \\
& & \times 
\left[ s^2 -d(d-1)\frac{1-\lambda^2}{4}\left(\frac{I_1(\varrho)}{I_0(\varrho)}\right)^2 
-\frac{d}{2}\frac{1-\lambda^2}{4}\left(1+\frac{I_2(\varrho)}{I_0(\varrho)}\right)  \right] \ .
\label{B9}
\EEA
Eqs.~(\ref{B9}) and (\ref{B8}) are eqs.~(\ref{2.11},\ref{2.12}) in the main text. 

Indeed, the dimension $d$ can now be considered as a real parameter, which offers obvious conceptual advantages. 
For $s\geq s_c=\frac{1+\lambda}{2}d$ and $\lambda\ne 0$, this integral is convergent for all $d>1$.
While this representation, as it stands, holds true for all values of $\lambda$, the asymptotic analysis will become more simple
for $0<\lambda<1$, where the possibility of zeroes of the $I_n(\varrho)$, with $n\geq 1$, need not be taken into account.

\appsection{C}{Asymptotic behaviour}

We analyse the spherical constraint (\ref{2.11}) and derive the asymptotic relations 
(\ref{gl:gtg}) for generic couplings $0<\lambda<1$. 

{\bf 1.} For $1<d<3$, the leading contribution to the shift $t_g$ in the coupling $g$ is non-analytic. 
Considering the spherical constraint (\ref{2.11}), non-analytic contributions 
come from large values of $u$ in one of the integrals. 
Combining eqs.~(\ref{2.11},\ref{2.15}), we must analyse 
\BEA 
t_g &:=& \sqrt{\frac{8J}{\hbar^2}}\left(\frac{1}{\sqrt{g}} -\frac{1}{\sqrt{g_c}}\right)
= \frac{1}{\pi s^{3/2}} \int_{\eta}^{\infty} \!\D u 
\int_0^1 \!\D x\: \frac{\exp\left[-us\right]}{\sqrt{x(1-x)}} \left[I_0(\varrho)\right]^d \times  
\\ \nonumber
& & \times \left[ s^2 -d(d-1)\frac{1-\lambda^2}{4}\left(\frac{I_1(\varrho)}{I_0(\varrho)}\right)^2 
-\frac{d}{2}\frac{1-\lambda^2}{4} \left(1+\frac{I_2(\varrho)}{I_0(\varrho)}\right)  \right] \ .
\EEA
for a spherical parameter $s=s_c+\sigma=\demi(1+\lambda)d + \sigma$ 
in the vicinity of $\sigma\to 0$. Here $\eta$ is a cut-off which helps to isolate
the non-analytic contributions to $t_g$ and we shall let $\eta\to\infty$ at the end. 
Because of (\ref{2.12}), the argument $\varrho$ of
the modified Bessel functions never vanishes for $0<\lambda<1$. Then, 
in order to obtain the leading behaviour in $\sigma$, it is enough to use the leading asymptotic behaviour 
$I_n(\varrho)\simeq e^{\varrho}/\sqrt{2\pi\varrho} \left( 1 + {\rm O}(1/\varrho)\right)$ \cite{Abra65} of the
modified Bessel functions. Then $I_n(\varrho)/I_0(\varrho)\simeq 1$ to leading order in $1/\varrho$ for $n=1,2$ and we arrive at 
\BEA
t_g \simeq \frac{1}{\pi\sqrt{s^3}} \int_{\eta}^{\infty} \!\D u \int_0^1 \!\D x\: \frac{\exp\left[-us\right]}{\sqrt{x(1-x)}} 
\left(\frac{\exp \varrho}{\sqrt{2\pi \varrho}}\right)^d \bigg\{  s^2 -d(d-1)\frac{1-\lambda^2}{4} -d\frac{1-\lambda^2}{4}  \bigg\} \ .
\EEA
For convenience,we recall the definition of $\varrho$ from (\ref{2.12}) 
\begin{equation} \label{C3}
\varrho = \varrho(u,x,\lambda) = u\left(x\frac{1+\lambda}{2}+(1-x)\frac{1-\lambda}{2}\right)
\end{equation}
and absorb into a single constant $\kappa$ several purely numerical factors 
\begin{equation}\label{eq:k1}
\kappa :=\left.\left(s^2 -d^2\frac{1-\lambda^2}{4}\right)\frac{\pi^{-\frac{d}{2}-1}}{s^{3/2}}\right|_{s=s_c} 
= \sqrt{\frac{d \lambda^2}{2(1+\lambda)\,}}\:\frac{1}{\pi^{1+d/2}} \ .
\end{equation}
such that the constraint becomes more compactly 
\BEA
t_g &\simeq& \kappa \int_{\eta}^{\infty} \!\D u \int_0^1 \!\D x\: \frac{\exp\left[-us\right]}{\sqrt{x(1-x)}} 
\frac{\exp\left[ ux\frac{1+\lambda}{2}d\right]
\exp\left[u(1-x)\frac{1-\lambda}{2}d\right]}{u^{d/2}\left(1-\lambda(1-2x)\right)^{d/2}} 
\nonumber \\
&=& \kappa \int_{\eta}^{\infty} \!\D u \int_0^1 \!\D x\: \frac{\exp\left[-u\sigma\right]}{\sqrt{x(1-x)}} 
\frac{\exp\left[ -u(1-x)d\lambda\right]}{u^{d/2}\left(1-\lambda(1-2x)\right)^{d/2}} \nonumber \\
&=& 2\kappa \int_{\eta}^{\infty} \!\D u\:  u^{-d/2}\exp\left[-u\sigma\right]\int_0^{\frac{\pi}{2}} \!\D\phi\:
\frac{\exp\left[ -ud\lambda\cos^2\phi\right]}{\left(1-\lambda\cos2\phi\right)^{\frac{d}{2}}} \nonumber \\
&=& 2\kappa \sigma^{\frac{d}{2}-1} \int_0^{\pi/2} \frac{\D\psi}{\left(1+\lambda\cos2\psi\right)^{d/2}}
\int_{\eta\sigma}^\infty \!\D v\:v^{-d/2}\exp\left[-v-vd\lambda\frac{\sin^2\psi}{\sigma}\right] \ .
\label{C5}
\EEA
where we used $s=\demi(1+\lambda)(x+1-x)d+\sigma$ in the 2$^{\rm nd}$ line and changed variables several times, 
in the 3$^{\rm rd}$ line according to $x=\sin^2\phi$, and in the 4$^{\rm th}$ line $v=u\sigma$ and 
$\phi = \frac{\pi}{2} - \psi$, and also used $\cos\phi=\sin\psi$ and $\cos 2\phi=-\cos 2\psi$.

We are interested in the asymptotic behaviour near criticality, when $0<\sigma \ll 1$. 
Furthermore the main contribution to the $\psi$--integral, for $v$ still finite, will come from 
the region where $\sin^2\psi /\sigma = {\rm O}(1)$. But in the $\sigma\to 0^+$ limit we consider 
here $\psi$ will be small as well so that we can replace $\sin \psi \simeq \psi$. Then the main contribution to 
this particular integral in (\ref{C5}) should come from the region
\begin{equation}
\psi^2 \lesssim \sigma \ .
\end{equation}
Hence the leading term can be obtained by replacing the upper limit in the $\psi$-integral in (\ref{C5}) by infinity. 
Changing the order of integrations, we find
\BEA 
t_g &\simeq& 2\kappa \sigma^{\frac{d}{2}-1}   
\int_{\eta \sigma}^\infty \frac{\D v}{v^{d/2}}\exp\left[-v\right] 
\int_0^{\infty} \frac{\D\psi}{ \left(1+\lambda\right)^{d/2}} \exp\left[-vd\lambda\frac{\psi^2}{\sigma}\right] 
\nonumber \\
&=& \frac{2\kappa}{(1+\lambda)^{d/2}}\, \sigma^{\frac{d}{2}-1}   
\int_{\eta \sigma}^\infty \frac{\D v}{v^{d/2}}\exp\left[-v\right] \sqrt{\frac{\sigma}{vd\lambda}} \frac{\sqrt{\pi}}{2} 
\nonumber \\ \label{eq:C7}
&=& \sigma^{(d-1)/2}\, \kappa \sqrt{\frac{\pi}{ d\lambda}}\left(1+\lambda\right)^{-d/2}\,\Gamma\left(\frac{1-d}{2},\eta\sigma\right)
\:=\: \sigma^{(d-1)/2}\, \frac{ \Gamma\left(\frac{1-d}{2},\eta\sigma\right)\sqrt{ \lambda/2}}{[\pi(1+\lambda)]^{(d+1)/2}}
\ 
\EEA
with the incomplete Gamma function $\Gamma(a,x)$ \cite{Abra65}.
Next, we have to carry out the two limiting processes, first $\sigma \rightarrow 0^+$ and then $\eta \rightarrow \infty$, in 
exactly this order. Defining the Gamma function via analytical continuation, for $1<d<3$ 
we simply have $\lim_{\sigma\to 0} \Gamma(\frac{1-d}{2},\sigma)=\Gamma(\frac{1-d}{2})$ and obtain 
\BEA
t_g \simeq \sigma^{(d-1)/2}\, \frac{ \Gamma\left(\frac{1-d}{2}\right)\sqrt{ \lambda/2}}{[\pi(1+\lambda)]^{(d+1)/2}} 
=: A_<\, \sigma^{(d-1)/2}
\mbox{\rm ~~~ for $1<d<3$}
\EEA

{\bf 2.} For $d=3$, we can repeat the analysis leading to (\ref{eq:C7}). 
However, the limit $\sigma\to 0^+$ in the incomplete Gamma function has
to be taken more carefully. Using \cite[eqs. (6.5.19, 5.1.11)]{Abra65}, one has a logarithmic term 
\BEA 
\Gamma(-1,x) \simeq \frac{1}{x}+C_E -1 +\ln(x)-\frac{x}{2}+{\rm O}(x^2)
\EEA 
where $C_E \approx 0.5772\ldots$ is  Euler's constant. 
Consequently, we find for the $\sigma$-dependence in $t_g$
\BEQ 
\sigma \Gamma(-1,\eta\sigma) \simeq \frac{1}{\eta}+\left[C_E -1 +\ln(\eta\sigma)\right]\sigma+{\rm O}(\sigma^2) 
\simeq \sigma \ln \sigma
\EEQ
In the last expression, we merely retain the most singular term when $\sigma\to 0^+$ 
with $\eta$ finite and then dropped those terms  which vanish in the 
$\eta\to\infty$ limit. The leading non-analytic contribution in (\ref{eq:C7}) is 
\BEA
t_g \simeq \frac{\sqrt{ \lambda/2}}{[\pi(1+\lambda)]^{2}}\:  \sigma \ln \sigma =: A_3 \, \sigma\ln \sigma 
\mbox{\rm ~;~~ for $d=3$}
\EEA

{\bf 3.} For $d>3$, the non-analytic contribution from eq.~(\ref{eq:C7}) 
$t_g \sim \sigma^{(d-1)/2}$ is of higher order than linear. 
The leading term in $t_g$ now comes from the the analytic contributions to (\ref{2.11}) 
which was previously subtracted from the left-hand side.
The leading correction term is found by a straightforward expansion in $\sigma$. We also introduce the short-hand
$F(d,\lambda,\rho):=
-d(d-1)\frac{1-\lambda^2}{4}\frac{I_1(\rho)^2}{I_0(\rho)^2}
-\frac{d}{2}\frac{1-\lambda^2}{4}\left(1+\frac{I_2(\rho)}{I_0(\rho)}\right)$
which is obviously independent of $s$. Hence, recalling also (\ref{C3})
\BEA \
\sqrt{\frac{8\pi^2J}{g\hbar^2}} &=& 
\int_0^\infty \!\D u \int_0^1\frac{\D x \ I_0(\varrho)^d}{\sqrt{x(1-x)}}\: 
s^{-3/2} \exp\left[-us\right]\left(s^2+F(d,\lambda,\varrho)\right)
\nonumber\\
&\simeq& \int_0^\infty \!\D u \int_0^1\frac{\D x \ I_0(\varrho)^d}{\sqrt{x(1-x)}}
\left[\sqrt{s_c}+\frac{1}{\sqrt{s_c^3}}-\left(\frac{2us_c-1}{2\sqrt{s_c}} + \frac{3+2us_c}{2s_c^{5/2}}
F(d,\lambda,\varrho) \right)\sigma\right] e^{-u s_c} 
\nonumber
\EEA 
In this expansion, the zeroth order gives $g_c$ and the first order 
gives the required linear contribution $t_g \simeq A_>\, \sigma$, 
where $A_>$ is given below in (\ref{C15}). Its value must be found numerically.

Summarising, we have found, for $0<\lambda\leq 1$
\BEQ \label{glC:gtg}
t_g \simeq \left\{ 
\begin{array}{ll} A_<\: \sigma^{(d-1)/2} & \mbox{\rm ~;~~ if $1<d<3$} \\
                  A_3\: \sigma \ln \sigma & \mbox{\rm ~;~~ if $d=3$} \\
		          A_>\: \sigma & \mbox{\rm ~;~~ if $d>3$}
\end{array} \right.
\EEQ
with the following constant amplitudes (derived here for $0<\lambda<1$ but which can be continued to $\lambda=1$ as well) 
\BEA
A_< &:=& \frac{ \Gamma\left(\frac{1-d}{2}\right)\sqrt{ \lambda/2}}{[\pi(1+\lambda)]^{(d+1)/2}}
\label{C13} \\
A_3 &:=& \frac{ \sqrt{ \lambda/2}}{[\pi(1+\lambda)]^{2}}
\label{C14} \\
A_> &:=& -\frac{1}{\pi}\int_0^\infty \!\D u \int_0^1\frac{\D x \ I_0(\varrho)^d}{\sqrt{x(1-x)}}
\left[\frac{2us_c-1}{2\sqrt{s_c}} + \frac{3+2us_c}{2s_c^{5/2}}\,
F(d,\lambda,\varrho)\right] e^{-u s_c} 
\label{C15}
\EEA
with $s_c = (1+\lambda)d/2$, $F(d,\lambda,\rho)$ was defined above and (\ref{C3}) 
was used. On the other hand, for $\lambda>1$ the argument $\varrho$ of the $I_n(\varrho)$, 
as given by (\ref{C3}), can vanish, the analysis leading to (\ref{eq:C7}) has to be re-done
and (\ref{glC:gtg}) cannot be expected to remain valid.  


\appsection{D}{Spin-spin correlator}

Using the representation (\ref{1.5}) in terms of ladder operators and then the canonical transformation (\ref{A4},\ref{A8}) from appendix~A, 
the spin-spin correlator is given by 
\BEA
\left< S_{\vec{n}} S_{\vec{m}}\right> &=&\sqrt{\frac{\hbar^2g}{8Js}}
\left< \left(\wht{a}_{\vec{n}}+\wht{a}_{\vec{n}}^\dagger\right) \left(\wht{a}_{\vec{m}}+\wht{a}_{\vec{m}}^\dagger\right)\right>\\
&=&\sqrt{\frac{\hbar^2g}{8Js}}\sum_{\vec{k}, \vec{k}'}(\vec{\Psi}_{\vec{k}})_{\vec{n}} (\vec{\Psi}_{\vec{k}'})_{\vec{m}}
\left[\left<\wht{b}_{\vec{k}} \wht{b}_{\vec{k}'} \right> +\left<\wht{b}_{\vec{k}}^\dagger \wht{b}_{\vec{k}'} \right>  + \mbox{h.c.}\right]
\EEA
Since the ladder operators are bosonic, they obey Bose-Einstein-statistics. Hence
\BEA
\left<\wht{b}_{\vec{k}} \wht{b}_{\vec{k}'} \right> =\left<\wht{b}_{\vec{k}}^\dagger \wht{b}_{\vec{k}'}^\dagger \right> = 0
\ ;\ \left<\wht{b}_{\vec{k}}^\dagger \wht{b}_{\vec{k}'} \right> = \delta_{\vec{k}, \vec{k}'} 
\left(\exp\left[\Lambda_{\vec{k}}/T\right]-1\right)^{-1}
\EEA
This immediately leads to 
\BEQ
\left< S_{\vec{n}} S_{\vec{m}}\right>= \sqrt{\frac{\hbar^2g}{8Js}}
\sum_{\vec{k}} (\vec{\Psi}_{\vec{k}})_{\vec{n}} (\vec{\Psi}_{\vec{k}})_{\vec{m}} \coth \left[\Lambda_{\vec{k}}/(2T) \right]
\EEQ
Using the real representation of the vector $\vec{\Psi}_{\vec{k}}$ from appendix~A, we find for the correlator in the continuum limit,
with $\vec{m}=\vec{n}+\vec{r}$
\BEQ \label{D5}
\left< S_{\vec{n}} S_{\vec{n}+\vec{r}}\right> = \sqrt{\frac{\hbar^2g}{8Js}}
\int_{\cal{B}}\frac{\D \vec{k}}{(2\pi)^d}\sqrt{\frac{2s-(1-\lambda)\sum_{j=1}^d \cos k_j}{2s-(1+\lambda)\sum_{j=1}^d \cos k_j}}
\coth \left[\Lambda_{\vec{k}}/(2T) \right] \prod_{j=1}^d \cos \left(r_j k_j\right)
\EEQ
and spatial translation-invariance is explicit, so that we can set $\vec{n}=\vec{0}$ from now on. 
Eq.~(\ref{D5}) is an exact expression for any temperature $T$. 

{\bf 1.} For $\lambda\ne0$, consider the quantum phase transition at $T=0$. Then (\ref{D5}) simplifies to 
\BEQ\label{gl:cf}
\left< S_{\vec{0}} S_{\vec{r}}\right> = 
\\ \sqrt{\frac{\hbar^2g}{8Js}}\int_{\cal{B}}\frac{\D \vec{k}}{(2\pi)^d}
\sqrt{\frac{2s-(1-\lambda)\sum_{j=1}^d \cos k_j}{2s-(1+\lambda)\sum_{j=1}^d \cos k_j}\,}\,
\prod_{j=1}^d \cos \left(r_j k_j\right)
\EEQ
Because of explicit rotation-invariance, we can choose axes such that $\vec{r}=(R,0\ldots,0)$.
Now, eq.~(\ref{gl:cf})
can be factorised by the same techniques as used in appendix~B to factorise the spherical constraint. We find, with $\varrho$ from eq. (\ref{C3})
\BEQ
\left< S_{0} S_{R}\right> = \sqrt{\frac{\hbar^2g}{8Js}}\frac{1}{\pi} \int_0^\infty \!\D u \int_0^1 \frac{\D x\exp[-us]}{\sqrt{x(1-x)}}
\left[s-\frac{1-\lambda}{2}\left((d-1)\frac{I_1(\varrho)}{I_0(\varrho)}+\frac{I_{R}'(\varrho)}{I_{R}(\varrho)}\right) \right]
I_0^{d-1}(\varrho)I_{R}(\varrho)
\EEQ

In order to work out the correlator from this representation, we now analyse the main contributions to the $u$-integral.
Since the integrand vanishes for $u=0$ and $u= \infty$, it will have a maximum at some intermediate value $u_{\rm max}$ 
and if the integrand is sufficiently peaked
around $u_{\rm max}$, this will give the main contribution. Now, the leading term of the series expansion 
$I_R(\rho) \simeq(\rho/2)^R/\Gamma(R+1)$, for small arguments $\rho\ll 1$, shows that for $u$ not too large, the integrand
will roughly behave as $u^{R}e^{-u}$ such that $u_{\rm max}\sim R$. Since we merely interested in the large-$R$ limit, it follows that
the contribution of small values of $u$ to the integral is negligible to leading order. Therefore, in order to estimate $\langle S_0 S_{R}\rangle$, 
it is enough to use the $\rho\gg 1$ asymptotic form $I_\nu(\rho) \simeq (2\pi \rho)^{-1/2}\exp\left[\rho-\frac{\nu^2}{2\rho}\right]$ 
of the Bessel functions, such that  
\BEQ
\left< S_0S_R\right> \simeq \sqrt{\frac{\hbar^2g}{8Js}}\frac{1}{\pi}\left(s-\frac{1-\lambda}{2}d\right)
\int_0^\infty\!\D u \int_0^1\D x\, \frac{(2\pi\varrho)^{-d/2}}{\sqrt{x(1-x)}}\,
\exp\left[ d\varrho -us - \frac{R^2}{2\varrho}\right]
\EEQ
This can be evaluated following the lines of appendix~C. We find
\BEQ
\left< S_0S_R\right> = \sqrt{\frac{\hbar^2g}{8Js}}\frac{(1+\lambda)^{-d/2}}{\pi^{(d+1)/2}}\frac{(s-\frac{1-\lambda}{2}d)}{\sqrt{\lambda d\,}\,}\:
\sigma^{(d-1)/2}\int_0^\infty \!\D v\ v^{-(d+1)/2}\exp\left[-v-\frac{R^2\sigma}{ (1+\lambda)}\frac{1}{v} \right]
\EEQ
This equation can be rewritten, using the identity 
$K_\nu(x)= \demi \left(\frac{x}{2}\right)^\nu \int_0^\infty \D v v^{-\nu-1}\exp\left[-v-x^2/(4v)\right]$ \cite{Grad07} 
for the modified Bessel function of the second kind, to obtain\footnote{For $\lambda =1$, eq.~(\ref{D10}) 
reproduces the well-known result \cite[eq.(13)]{Oliv06}, 
if one takes into account that because of the normalisation $\left< \sum_{\vec{n}} S_{\vec{n}}^2\right> = {\cal{N}}/4$ chosen
in \cite{Oliv06}, one must renormalise $s\mapsto s/4$ to ensure matching pre-factors.}
\BEQ \label{D10}
\left< S_0S_R\right> =
\sqrt{\frac{\hbar^2g}{Js}} \frac{2^{-d/2}}{\pi^{(d+1)/2}}\frac{s-\frac{1-\lambda}{2}d}{\sqrt{\lambda(1+\lambda) d}\,}
\: \left(\frac{1}{\xi R}\right)^{(d-1)/2} K_{\frac{d-1}{2}} \left(\frac{R}{\xi}\right)
\EEQ
and where the correlation length was identified as $\xi :=\frac{1}{2}\sqrt{\frac{1+\lambda}{\sigma}}$. 
Very close to criticality, $\xi$ diverges, hence $R/\xi\ll 1$. At some finite distance from $g_c$, one has on the contrary $R/\xi\gg 1$. 
Now, using the leading expansions \cite[eqs. (9.6.9,9.7.2)]{Abra65}, one has the asymptotic behaviour 
\BEQ
\left< S_0 S_R\right> \simeq \sqrt{\frac{\hbar^2 g}{ J s}\,}\,\frac{s-\frac{1-\lambda}{2}d}{\pi^{(1+d)/2}\sqrt{\lambda(1+\lambda)d}} \times \left\{
\begin{array}{ll} 
2^{-3/2} \ \Gamma\left(\frac{d-1}{2}\right)\cdot R^{1-d} & \mbox{\rm ~;~~ if $R\ll \xi$} \\[0.10cm]
2^{-d/2}\sqrt{\pi/2} \cdot \xi^{1-d}\, (\xi/R)^{d/2}\, \	e^{-R/\xi}  & \mbox{\rm ~;~~ if $R\gg \xi$}
\end{array} \right.
\EEQ
with $s=s_c+\sigma$, $\sigma$ is related to $\xi$ and the value of $g$ has to be taken from the spherical constraint. 

{\bf 2.} If $\lambda = 0$, we have to do a more careful analysis, since the zero-temperature contribution is completely disconnected.
From eq.~(\ref{gl:cf}), we see a $\delta_{R,0}$ contribution arising. Thus, the leading non-trivial contributions in this particular case are thermal and
we have to re-investigate the correlation function for non-zero temperatures. Hence we return to eq.~(\ref{D5}), 
as well as to (\ref{gl:2.5}) for the spherical constraint, in order to find the thermal corrections to the critical coupling constant $g_c$. 
Since we are still interested in a certain low-temperature limit and not in the thermal
transition, we take $0<T\ll 1$ and use the asymptotic expansion $\coth x \simeq 1+2\exp(-2x)$ 
to obtain the leading correction. The spherical constraint in zero field then reads 
\BEQ
1 = \sqrt{\frac{\hbar g}{8Js}} + 2\sqrt{\frac{\hbar^2 g}{8Js}}\exp\left( -2zs\right)
I_0\left(z \right)^d\left[1+\frac{d}{2s}\frac{I_1\left(z\right)}{I_0\left(z\right)}\right]
\EEQ
with the argument $z:=\sqrt{{gJ\hbar^2}/{2T^2s}}$. In the low-temperature limit, $z\rightarrow \infty$. From the 
asymptotic expansion of the modified Bessel functions, we find
\BEQ
\sqrt{\frac{8Js}{\hbar^2 g}} \simeq 1 + \frac{2\e^{-2\sigma z}}{(2\pi z)^{d/2}} \left[ 1 +\frac{d}{2s}\left(1-\frac{1}{2z}\right)\right]
\EEQ
Studying this equation up to the leading order in $1/z$, at the quantum critical point $\sigma = 0$, we deduce the implicit 
equation for the critical coupling constant $g_c=g_c(0,d;T)$
\BEQ\label{gl:gctherm}
\sqrt{\frac{J}{\hbar^2g_c}} = \sqrt{\frac{1}{4d}} +\frac{d^{-1/2}}{(2\pi)^{d/2}}\left(\frac{d}{g_cJ\hbar^2} \right)^{d/4} T^{d/2}
\EEQ
First of all we see, that this equation is consistent with the zero-temperature limit and reproduces $g_c(0,d;0)=4dJ/\hbar^2$ correctly. 
While for $d=2$, there is a simple closed solution
\BEQ
g_c(0,2;T) = \left(\sqrt{\frac{8J}{\hbar^2}} - \sqrt{\frac{2}{\pi^2}} \frac{T}{\sqrt{J\hbar^2}}\right)^2
\EEQ
eq.~(\ref{gl:gctherm}) cannot be solved in closed form in general. 

For large distances, the same techniques as before, applied to (\ref{D5}), lead for $\lambda=0$ to 
\BEQ
\langle S_0 S_R\rangle = \sqrt{\frac{\hbar^2g}{8Js}} + \sqrt{\frac{\hbar^2g}{2Js}}\exp\left(-2zs\right) I_0(z)^{d-1}I_R(z)
\EEQ
Using the asymptotic expansion for the Bessel functions, we find at the critical point $g=g_c$
\BEQ \label{D17}
\langle S_0 S_R\rangle = \sqrt{\frac{\hbar^2 g_c}{4dJ}}\: \delta_{R,0} + \sqrt{\frac{\hbar^2g_c}{dJ}}\left(\frac{T^2d}{4\pi^2g_cJ\hbar^2} \right)^{d/4} \exp\left( - \frac{R^2T}{2}\sqrt{\frac{d}{g_cJ\hbar^2}} \right)
\EEQ
\appsection{E}{Critical coupling $g_c(\lambda,d)$ close to $\lambda = 0$}

In order to prove (\ref{2.24}) and to understand the unexpected behaviour of the function $g_c(\lambda,d)$ close to $\lambda =0$, we re-investigate 
the equation (recall the definition (\ref{B8}) of $\varrho=\varrho(u,\lambda,d)$) 
\BEA\nonumber
\sqrt{\frac{J\pi^2(1+\lambda)^3d^3}{\hbar^2g_c(\lambda,d)}} &=& 
\int_0^\infty \!\D u \int_0^1 \!\D x\, \frac{\exp\left[-u\frac{1+\lambda}{2}d\right]}{\sqrt{x(1-x)}} 
\left[I_0(\varrho)\right]^d \times \left\{ 
\left(\frac{1+\lambda}{2}d\right)^2 \right.\\ 
& & \left.-d(d-1)\frac{1-\lambda^2}{4}\left(\frac{I_1(\varrho)}{I_0(\varrho)}\right)^2 
-\frac{d}{2}\frac{1-\lambda^2}{4}\left(1+\frac{I_2(\varrho)}{I_0(\varrho)}\right)  \right\} 
\label{E1} \\
&=& \sqrt{\frac{J\pi^2d^3}{\hbar^2g_c(0,d)}} + G^{(1)} + G^{(2)}
\nonumber
\EEA
where the two contributions $G^{(1)}$ and $G^{(2)}$ describe the leading behaviour 
in $\lambda$, which are non-analytic and analytic, respectively. 

First, we consider the case $1<d<2$, when the  leading behaviour is given by the non-analytic term $G^{(1)}$. 
After a change of variable $\varrho  = \frac{u}{2}\left(1-\lambda + 2\lambda x \right)$ in (\ref{E1}), 
we divide the $\varrho$-integral in two parts $\int_0^\infty \D \varrho  = \int_0^\eta \D\varrho+ \int_\eta^\infty\D \varrho$. 
In the limit $\lambda \rightarrow 0^+$ and $\eta \rightarrow \infty$, 
the first integral  reduces to $g_c(0,d)$ while the second integral will give the desired non-analytic term $G^{(1)}$, for small $\lambda$.
As in appendix~C, $G^{(1)}$ is analysed via the asymptotic expansions of the Bessel functions \cite{Abra65},  which gives 
\BEA
G^{(1)} &=& \int_\eta^\infty \!\D \varrho \int_0^1 \frac{\D x}{\sqrt{x(1-x)}}
\frac{\exp\left[-2d\varrho\lambda \frac{1-x}{1-\lambda+2\lambda x }\right]}{(1-\lambda+2\lambda x)(2\pi\varrho)^{d/2}}d^2\lambda
\nonumber \\
&=& 2d^2\lambda^{d/2}\int_0^{\pi/2}\D \vartheta\int_{\lambda \eta}^\infty \!\D y\:
\frac{\exp\left[-2dy \sin^2 \vartheta \right]}{(2\pi y)^{d/2}}
\nonumber \\
&=& 2d^2\lambda^{d/2}\int_{0}^\infty \!\D y\:
\exp\left[-dy\right]I_0(yd)(2\pi y)^{-d/2}
\EEA
where in the second line, we made the substitutions $y= \lambda \varrho$ and $x=\cos^2\vartheta$ and in the third line recalled the identity
$\sin^2\vartheta = \demi(1-\cos 2\vartheta)$ to derive $\int_0^\infty\D \vartheta\, e^{-A\sin^2\vartheta}= \frac{\pi}{2}\,e^{-A/2}I_0(A/2)$ from the 
defining integral representation of $I_0(x)$ \cite{Abra65}. 
For $1<d<2$, this is indeed the leading contribution. Explicitly, using \cite[eq. (2.15.3.3)]{Prudnikov2}, this further simplifies to 
\BEQ \label{E3}
\sqrt{\frac{J}{\hbar^2}}g_c(\lambda,d)^{-1/2} \simeq \frac{1}{\sqrt{4d}} 
+\frac{d^{(d-1)/2}}{2\pi^{(d+1)/2}}\frac{\Gamma(1-d/2)\Gamma((d-1)/2)}{\Gamma(d/2)}\:\lambda^{d/2} 
\EEQ 
with a finite, positive amplitude for all dimensions $1<d<2$. 

For $d>2$, the non-analytic contribution $G^{(1)}$ in (\ref{E3}), analytically continued in  $d$,  
is dominated by a new analytic contribution $G^{(2)}$. To obtain this, one must formally expand the integrand in (\ref{E1})
to first order in $\lambda$. Of course, such as a formal expansion is only admissible up to the order where the expansion coefficient(s) converge(s). 
Because of the definition (\ref{B8}) of $\varrho$, in principle the Bessel functions $I_n(\varrho)$ should be expanded
around $\lambda=0$. However, the leading term will introduce a factor $1-2x$ into the integrand and all these contributions vanish because of
$\int_0^1\!\D x\, (1-2x)/\sqrt{x(1-x)}=0$. Therefore, the additional contribution reads
\BEQ
G^{(2)} = -\lambda \frac{\pi d}{2} \int_0^{\infty} \!\D\varrho\: \varrho\, e^{-d\varrho} I_0^d(\varrho)
\left[ d^2\left(1 -\frac{I_1^2(\varrho)}{I_0^2(\varrho)}\right) 
+\frac{d}{2}\frac{2 I_1^2(\varrho)-I_0^2(\varrho)-I_0(\varrho)I_2(\varrho)}{I_0^2(\varrho)}
\right] + {\rm O}(\lambda^2)
\EEQ
and the integral over $x$ has become trivial in the $\lambda\to 0$ limit. 
This contribution is linear in $\lambda$ and hence will dominate over $G^{(1)}$ for $d>2$. 
In order to study its convergence, we split as usual $\int_0^{\infty}\!\D\varrho = \int_0^{\eta} \!\D\varrho + \int_{\eta}^{\infty} \!\D\varrho$
and analyse the convergence of the second integral. Using the asymptotic expansion
of the $I_n(\varrho)$ up to next-to-leading term in $1/\varrho$ \cite{Abra65}, the large-$\eta$ behaviour of $G^{(2)}$ is given by
$-\frac{\lambda d}{2}(2\pi)^{-d/2} \int_{\eta}^{\infty} \!\D\varrho\: \varrho^{-d/2}$ and this converges for $d>2$. For $d<2$ however,
the integral $G^{(2)}$ diverges such that the formal expansion used to derive it does not exist. Then (\ref{E3}) gives indeed the leading contribution
to $g_c(\lambda,d)$ for $\lambda\ll 1$. 

This proves (\ref{2.24}) in the main text. 

%

\noindent 
{\bf Acknowledgements:} It is a pleasure to thank J.-Y. Fortin, G. Morigi and A. Pikovsky for useful discussions. 
Part of this work was done during the workshop ``Advances in Non-equilibrium Statistical Mechanics''. 
MH gratefully thanks the organisers and the Galileo Galilei Institute for Theoretical Physics for their warm 
and generous hospitality and the INFN for partial support. This work 
was also partly supported by the Coll\`ege Doctoral franco-allemand Nancy-Leipzig-Coventry
({\it `Syst\`emes complexes \`a l'\'equilibre et hors \'equilibre'}) of UFA-DFH. 
SW is grateful to UFA-DFH for financial support through grant CT-42-14-II. 


\end{document}